\documentclass[preprint]{aa}

\usepackage{graphicx}

\def\msol{\hbox{\kern 0.20em $M_\odot$}}
\def\lsol{\hbox{\kern 0.20em $L_\odot$}}
\def\rsol{\hbox{\kern 0.20em $R_\odot$}}
\def\sr{\hbox{\kern 0.20em sr}}
\def\srmu{\hbox{\kern 0.20em sr$^{-1}$}}

\def\g{\hbox{\kern 0.20em g}}
\def\gmu{\hbox{\kern 0.20em g$^{-1}$}}
\def\kg{\hbox{\kern 0.20em kg}}
\def\pc{\hbox{\kern 0.20em pc}}

\def\mum{\hbox{\kern 0.20em $\mu$m}}
\def\mumd{\hbox{\kern 0.20em $\mu$m$^{-2}$}}
\def\cm{\hbox{\kern 0.20em cm}}
\def\m{\hbox{\kern 0.20em m}}
\def\km{\hbox{\kern 0.20em km}}
\def\nm{\hbox{\kern 0.20em nm}}

\def\s{\hbox{\kern 0.20em s}}
\def\h{\hbox{\kern 0.20em h}}
\def\sec{\hbox{\kern 0.20em sec}}
\def\min{\hbox {\kern 0.20em min}}
\def\smu{\hbox{\kern 0.20em s$^{-1}$}}
\def\smd{\hbox{\kern 0.20em s$^{-2}$}}
\def\an{\hbox{\kern 0.20em an}}
\def\anmu{\hbox{\kern 0.20em an$^{-1}$}}
\def\deg{\hbox{\kern 0.20em $^{\rm o}$}}
\def\yr{\hbox{\kern 0.20em yr}}
\def\yrmu{\hbox{\kern 0.20em yr$^{-1}$}}
\def\Myr{\hbox{\kern 0.20em Myr}}
\def\Mymu{\hbox{\kern 0.20em Myr$^{-1}$}}
\def\K{\hbox{\kern 0.20em K}}
\def\pcmu{\hbox{\kern 0.20em pc$^{-1}$}}
\def\pcmd{\hbox{\kern 0.20em pc$^{-2}$}}
\def\pcmt{\hbox{\kern 0.20em pc$^{-3}$}}
\def\kms{\hbox{\kern 0.20em km\kern 0.20em s$^{-1}$}}
\def\kmpd{\hbox{\kern 0.20em km$^{2}$}}
\def\kpc{\hbox{\kern 0.20em kpc}}
\def\cms{\hbox{\kern 0.20em cm\kern 0.20em s$^{-1}$}}
\def\erg{\hbox{\kern 0.20em erg}}
\def\ergs{\hbox{\kern 0.20em erg}}
\def\cmpd{\hbox{\kern 0.20em cm$^2$}}
\def\cmmd{\hbox{\kern 0.20em cm$^{-2}$}}
\def\cmms{\hbox{\kern 0.20em cm$^{-6}$}}
\def\cmpt{\hbox{\kern 0.20em cm$^3$}}
\def\cmmt{\hbox{\kern 0.20em cm$^{-3}$}}
\def\mpd{\hbox{\kern 0.20em m$^2$}}
\def\mmd{\hbox{\kern 0.20em m$^{-2}$}}
\def\mpt{\hbox{\kern 0.20em m$^3$}}
\def\mmt{\hbox{\kern 0.20em m$^{-3}$}}
\def\mujy{\hbox{\kern 0.20em $\mu$Jy}}
\def\mjy{\hbox{\kern 0.20em mJy}}
\def\Mj{\hbox{\kern 0.20em MJy}}
\def\jy{\hbox{\kern 0.20em Jy}}
\def\ghz{\hbox{\kern 0.20em GHz}}
\def\srmd{\hbox{\kern 0.20em sr$^{-1}$}}

\def \mum{$\mu$m}

\def\G{\hbox{\kern 0.20em G}}

\def\twco{\hbox{${}^{12}$CO}}

\def\thco{\hbox{${}^{13}$CO}}

\def\ceio{\hbox{C${}^{18}$O}}

\def\htwo{\hbox{H${}_2$}}
\def\h13cop{\hbox{H$^{13}$CO$^{+}$}}

\def\hcop{\hbox{HCO$^{+}$}}

\newcommand{\Hp}{\hbox{H{\small II}}}

\newcommand{\jonetozero}{\hbox{$J=1\rightarrow 0$}}
\newcommand{\jtwotoone}{\hbox{$J=2\rightarrow 1$}}
\newcommand{\jthreetotwo}{\hbox{$J=3\rightarrow 2$}}

\begin{document}
\title{Star formation in the Trifid Nebula}
\subtitle{Cores and filaments}
\author{
B.~Lefloch
  \inst{1}
\and J.~Cernicharo
  \inst{2,1}
\and J.R.~ Pardo\inst{2} }

\offprints{lefloch@obs.ujf-grenoble.fr}

\institute{
  Laboratoire d'Astrophysique de l'Observatoire de Grenoble, 414 rue de la Piscine,
  38041 Grenoble Cedex, France \\
  e-mail: lefloch@obs.ujf-grenoble.fr
  \and
  DAMIR, CSIC, C./Serrano 121, E-28006 Madrid
  }

\abstract{} {We aim to characterize the properties of the prestellar and
protostellar condensations to  understand the star formation processes at work
in a young HII region} {We have obtained maps of the 1.25mm thermal dust
emission and the molecular gas emission  over a region of $20\arcmin \times 10
\arcmin$ around the Trifid Nebula (M20), with the IRAM 30m and the CSO
telescopes as well as in the mid-infrared wavelength with ISO and SPITZER. Our
survey is sensitive to features down to $\rm N(\htwo) \sim 10^{22}\cmmd$ in
column density.} {The cloud material is distributed in fragmented dense gas
filaments ($\rm n(\htwo)$ of a few $10^3\cmmt$) with sizes ranging from  1 to
$10\pc$. A massive filament, WF, with properties typical of Infra Red Dark
Clouds, connects M20 to the W28 supernova remnant. We find that these filaments
pre-exist the formation of the Trifid and were originally self-gravitating. The
fragments produced are very massive (typically 100\msol\ or more) and are the
progenitors of the cometary globules observed at the border of the \Hp\ region.
We could identify 33 cores, 16 of which are currently forming stars. Most of
the starless cores have typical \htwo\ densities of a few $10^4\cmmt$. They are
usually gravitationally unbound and have low masses of a few $\msol$.  The
densest starless cores (several $10^5\cmmt$) are located in condensation TC0,
currently hit by the ionization front, and  may be the site for the next
generation of stars. The physical gas and dust properties of the cometary
globules  have been studied in detail and have been found very similar. They
all are forming stars. Several intermediate-mass protostars have been detected
in the cometary globules and in the deeply embedded cores. Evidence of
clustering has been found in the shocked massive cores TC3-TC4-TC5.} {M20 is a
good example of massive-star forming region in a turbulent, filamentary
molecular cloud. Photoionization appears to play a minor role in the formation
of the cores. The observed fragmentation is well explained by MHD-driven
instabilities and is usually not related to M20. We propose that the nearby
supernova remnant W28 could have triggered the formation of protostellar
clusters in nearby dense cores of the Trifid.}


\maketitle

\section{Introduction}

Many observations suggest that H{\small II} regions could play an important
role in spreading star formation throughout the Galaxy. As reviewed in
Elmegreen (2002), most of the embedded young stellar clusters in the solar
neighbourhood are adjacent to H{\small II} regions excited by more evolved
stars~: Orion, Perseus OB2, the Rosette nebula, W3-5, M17, to name but a few.
Yamaguchi et al. (1999) estimate that 10\%-30\% of stars in mass in the Galaxy
are formed under the influence of adjacent H{\small II} regions. In particular,
the condensations encountered at the border of the HII regions, the so-called
bright-rimmed Globules and cometary globules (CGs), have long been identified
as stellar nurseries or ``incubators'' (Sugitani et al. 1991,1994; Lefloch et
al. 1997; Dobashi et al. 2001). These condensations are local clumps which
emerge from the expanding molecular layer surrounding the nebula. Several
theoretical and observational studies  have shown the important role played by
the Ly-continuum radiation of the nebula in the evolution of these
condensations, which is now well understood (Bertoldi, 1989; Lefloch \&
Lazareff 1994, 1995). Systematic studies show that these ``stellar factories''
tend to form more massive and luminous (proto)stars, with luminosities $\geq
10^3\lsol$ (Sugitani et al. 1991, Dobashi 2001). It is estimated  that more
than 15\% of the massive stars form in CGs. The question of whether the
mechanical and radiative luminosities of OB stars can trigger the formation of
subsequent generations of stars, in particular of massive stars, in these
condensations is still an open question.

The ``collect and collapse'' model, first proposed by
Elmegreen \& Lada (1977), is particularly interesting as it enables the
formation of massive condensations and subsequently of massive objects out
of an initially uniform medium. Systematic studies
 have shown that this mechanism was at work
in some isolated Galactic \Hp\ regions and most likely responsible for the
massive YSOs observed at the border of the nebulae (Deharveng et al. 2003,2006;
Zavagno et al. 2007). In the case of a turbulent environment, the molecular gas
is no longer homogeneous but distributed in clumpy and filamentary structures;
it is not clear however what are the most efficient processes that account for
the formation of the subsequent generations of stars. Several groups have
investigated the impact of \Hp\ regions on the second-generation star
formation, and proposed two main mechanisms of induced star formation, either
direct triggering from \Hp\ regions and other pressure mechanisms associated
with hot stars, or gravity-driven streaming of gas along the filaments to build
dense cores (Nakamura, Hanawa \& Nakano, 1993; Tomisaka, 1995; Fiege \& Pudritz
2000; Fukuda \& Hanawa, 2000).

We have started a comprehensive study of a young \Hp\ region, the Trifid Nebula
(M20), small enough to allow a complete census of the young stellar and
protostellar population, to address this question. The Trifid Nebula appears as
a small dusty nebula in a young evolutionary age, at an heliocentric distance
of $1.68\kpc$ (Lynds et al. 1985). Mapping of the thermal dust and molecular
gas emission of the nebula at millimeter wavelengths revealed the presence of
several massive protostellar cores (dubbed TC1 to TC4), some of which are
exposed to the ionizing radiation of the exciting star (Cernicharo et al. 1998,
CL98). ISOCAM observations confirmed the extreme youth of the protostellar
sources detected in the cores TC3 and TC4 (see also Lefloch \& Cernicharo 2000,
hereafter LC00). So far, only the cometary globule TC2 has been analyzed in
detail  from its line and continuum emission at mid-infrared, millimeter and
centimeter wavelengths (Lefloch et al. 2002, herafter L02); its protostellar
content could not be characterized from the ISOCAM data, however.

The last phases of protostellar evolution in the nebula have been identified
thanks to ISOCAM and SPITZER. Deep images obtained by SPITZER in the range
$3.5-24\mu m$ with IRAC and MIPS have revealed the population of low-mass
objects which formed together with the exciting star of the nebula (Rho et al.
2006). In addition to 32 protostellar candidates, about 120 young stellar
objects (Class II) have been detected, which are surrounded by warm dust,
presumably distributed in protoplanetary disks. Analyzing the mid-infrared and
centimeter emission of the central stellar cluster members, we found that their
disks (components B, C, D) undergo copious photoionization (Lefloch et al.
2001, hereafter L01; Yusef-Zadeh et al. 2005).

In this article, we report on a large-scale survey of the cold dust emission
around the Trifid Nebula, complemented with molecular line emission data. Our
systematic mapping has allowed us to determine the evolutionary status of the
SPITZER Class~0/I candidates; it reveals the presence of stellar incubators in
the dense molecular material surrounding the \Hp\ region, and the progenitors
of the protostellar cores. This allows us to draw a full evolutionary sequence
for the incubators in the Trifid Nebula, from the early stage when they are
embedded in low-density material of the parental cloud, as illustrated  by the
massive pre-Orion cores TC3-TC4, to the late stage when the parental cocoon is
fully dissipated, leaving the protoplanetary disks exposed to the ionizing
radiation of the O star, as observed by L01 and Yusef-Zadeh et al (2005).

The article is organized as follows~: after presenting the observational data
(Sect.~2), we summarize the main results of the continuum survey (Sect.~3). We
then present in Sects.~4-5-6 detailed studies of typical cores which have
formed in the Western filament(TC0), the Eastern lane (TC1) and in the South of
the nebula (TC2 and TC9). We discuss in Sect.~7 the implications on the star
formation history of the Trifid. The main conclusions are summarized in
Sect.~8.

\section{The Observations}

\begin{figure*}
  \begin{center}
    \leavevmode
\caption{ {\em (left)}~Mid-infrared continuum emission at $8\mu m$ around M20
as observed with SPITZER (Rho et al. 2006). The dotted line traces the limit of
the field observed with the 1.3mm bolometer array. {\em (right)}~ Thermal dust
emission at $1300\mu m$ around M20. Contour levels are 5, 10, 60, 120, 240, 480
$mJy/11\arcsec$ beam. The location of the Class0/I sources identified by
SPITZER is marked by squares. } \label{iramsurvey}
\end{center}
\end{figure*}

\subsection{Continuum observations}

The ISO observations of the nebula are fully described in L02. Spitzer
observations are presented and discussed in Rho et al. (2006). In all maps
presented here the coordinates are given in arcsec offsets with respect to the
position of HD 164429A, the O star exciting the nebula~: $\rm \alpha= 18^h
02^m23.55^s$, $\rm \delta= -23^{\circ}01\arcmin51\arcsec$ (J2000.0).

Observations of the 1.25mm continuum emission in the Trifid were carried out in
March 1996 and March 1997 at the IRAM 30m-telescope (Pico Veleta, Spain) using
the MPIfR 19-channel bolometer array. The passband of the bolometer has an
equivalent bandwidth $\simeq 70$~GHz and is centered at 240 GHz. The beam size
of the telescope is $11\arcsec$. The final map was obtained by combining
several individual fields centered on the brightest condensations of the
nebula. Each field was scanned in the horizontal direction by moving the
telescope at a speed of $4\arcsec$ per sec; subsequent scans  are displaced by
$4\arcsec$. We used a chopping secondary at 2Hz with a throw of 30 to
$90\arcsec$ depending on the structure of the region to be mapped. The
dual-beam maps were reduced with the IRAM software for bolometer-array data
(Broguiere, Neri \& Sievers, 1995), which uses the EKH restoration algorithm
(Emerson, Klein \& Haslam, 1979). Zero-order baselines were subtracted from the
data prior to restoration. A few maps suffered from strong skynoise and were
processed using a noise-reduction algorithm that removes a correlated noise
component from each channel, based on the signal seen by the other channels of
the array (Broguiere et al. 1995). Since the 1.3mm continuum map was obtained
by combining several smaller maps taken in the dual-beam mapping mode, it is
insensitive to structures more extended than the typical size of an individual
bolometer map in azimuth, $\sim 4.5'$. The possibility of flux loss from
dual-beam filtering is inherent with the observational method and cannot be
discarded a priori for extended structures. This point was investigated in
detail by Motte et al. (1998), who found a negligible impact on compact
continuum cores. We note  a good agreement between the determinations of column
density based on dust and molecular gas, also in warm gas regions where
molecular gas depletion is known to be negligible (see e.g. Bacmann et al.
2002). This makes us confident that the flux losses resulting from dual-beam
filtering are rather small, except perhaps in the diffuse cloud in the TC15
region.

Calibration was checked against Mars and secondary calibrators at the IRAM 30m
telescope. The weather conditions were good and rather stable during the two
observing sessions. The opacity was monitored every hour on average and we
found typical zenith opacities between 0.1 and 0.35. Pointing was checked every
hour as the source transits at low elevation at Pico Veleta and corrections
were always found to be lower than $3\arcsec$. The relative calibration was
found to be good to within $10\%$ and the absolute uncertainty is estimated to
be about $15\%$.  The final map has a size of approximately $20\arcmin \times
10\arcmin$ and the rms is $8\mjy/11\arcsec$~beam over the whole map. Hence, it
is sensitive enough to detect at the $3\sigma$ level gas+dust column densities
of $\sim 10^{22}\cmmd$ per telescope beam.

In order to outline the weak extended dust components in the nebula, we
convolved our map with a Gaussian of $20\arcsec$ HPBW. The size of the
condensations is estimated taking into account the deconvolution from the main
beam and assuming a Gaussian flux distribution. All the results quoted in this
paper are based on the non-degraded map with $6\arcsec$ resolving power. The
continuum source properties are derived adopting a dust temperature $\rm T=
22\K$, a spectral index $\beta= 2$ and a dust opacity coefficient
$\kappa_{250}= 0.1\cmpd\gmu$. These parameters were obtained from the analysis
of the dust emission in the cores TC2 and TC4 (LC00, L02).

\subsection{Millimeter line observations}

\begin{table*}
\begin{flushleft}
\caption[]{Millimeter Line Observations. }
\begin{tabular}{|l|c|c|c|c|c|} \hline
Line & Frequency & Telescope & Beamwidth & $\rm B_{eff}$ & $\rm F_{eff}$ \\
     &  (GHz)    &           & (arcsec)  &               &         \\ \hline
$\hcop \jonetozero$ & 89.18852 & IRAM & 28 & 0.77 & 0.90 \\
CS $\jtwotoone$  & 97.98097  & IRAM  & 24 & 0.71  & 0.90 \\
                 &           & SEST  & 48 & 0.72  & 1.0 \\
$\ceio$ $\jonetozero$ & 109.28218 & IRAM &  22   &  0.68  & 0.92 \\
$\thco$ $\jonetozero$ & 110.20135 & IRAM &  22   & 0.68  & 0.92 \\
$\twco$ $\jonetozero$ & 115.27120 & IRAM  & 21    & 0.67  & 0.92 \\
CS $\jthreetotwo$ & 146.96905 & IRAM & 16 & 0.53  & 0.90 \\
                  &           & SEST & 32 & 0.66 & 1.0 \\
CO $\jtwotoone$ & 230.53800   & IRAM & 11 & 0.45 & 0.86 \\
CO $\jthreetotwo$ & 345.79599 & CSO & 22 & 0.75 & 1.0 \\
\hline
\end{tabular}
\end{flushleft}

\end{table*}

A region of 10 arcmin size centered on the  Trifid was mapped at full sampling
in the CO $\jtwotoone$, $\jonetozero$, $\thco$ $\jonetozero$ and $\ceio$
$\jonetozero$ lines with the IRAM 30m telescope (Pico Veleta, Spain) in July
1996, using the ``On-the-Fly'' technique. We used both and autocorrelator and
1MHz filterbanks as spectrometers. The autocorrelator provided a kinematic
resolution of $0.2\kms$ for all the transitions. The kinematic resolution
provided by the filterbanks ranges from $3.5\kms$ for the \jonetozero\
transitions of the CO isotopomers to $1.3\kms$ for the \jtwotoone\ transitions.
An ``intermediate''  reference position was taken 10 arcmin East of the Trifid;
the molecular emission at this position  was later observed in position
switching mode, adopting an Off position checked to be free of emission.

A large map of 0.5 deg $\times$ 0.5 deg was obtained at the CSO in the CO
\jthreetotwo\ transition during various observing runs between 1998 and 1999,
also using the "on-the-fly" technique with a reference position free of
emission. The map was sampled at $20\arcsec$, corresponding to approximately
one beamwidth. The receiver was connected to an AOS which provided a kinematic
resolution of $0.4\kms$.

Observations of the Trifid in the lines of CO $\jonetozero$, CS \jtwotoone\ and
\jthreetotwo\, and $\hcop \jonetozero$ were obtained with the SEST telescope
(La Silla, Chile) in September 1996 and October 1997. The observations were
carried out in position switching mode, using the same reference position as
for the OTF IRAM data. We observed a total area of $20' \times 20'$ with a
spacing of $20\arcsec$ over most of the region. Two strips of $2'\times 15'$ at
the eastern and western border of the field, where the molecular emission is
much less, were mapped with a $40\arcsec$ sampling. An AOS was used as a
spectrometer, which provided a spectral resolution of 43 kHz ($0.11\kms$ at the
frequency of the CO $\jonetozero$ line).

We have adopted the main-beam brightness temperature scale to express the
brightness of the lines studied here, with the exception of the \twco\ and
\thco\ lines. As it appears in our observations, bright $^{12}$CO and $^{13}$CO
emission extends over regions of several tens of arcmin.  At such scales, the
main-beam temperature is no longer a good approximation to the actual line
brightness because the error beam contribution is no longer negligible. The
antenna temperature corrected for rearward losses $\rm T_A^{*}$ appears as a
better estimate and we adopt this scale for the main isotopologue transitions
of CO. The validity of this approximation was discussed previously by L02 (see
Sect.~7.1). The observed molecular lines, their frequency, the telescope
beamwidth, the main-beam efficiency $\rm B_{eff}$ and the forward efficiency
$\rm F_{eff}$ are summarized in Table~1.

The physical conditions in the molecular gas associated with the condensations
were derived from the millimeter transitions of CS and the CO isotopomers. The
kinetic temperature was estimated from the CO \jtwotoone\ antenna temperature,
and found to lie in the range $25-30\K$. In some cores, a simple LTE analysis
of the \thco\ and \ceio\ lines could be performed and showed very consistent
results with the values derived from the \twco\ data. In all the condensations,
the ratio of the $\thco$ $\jonetozero$ to the $\ceio$ $\jonetozero$ intensities
was found to be very close to the canonical abundance ratio (8), within the
uncertainties. It implies that the \ceio\ line is optically thin, and the
$\thco$ either optically thin or moderately optically thick ($\tau \simeq 1$).
The very good consistency  of the results from  the different isotopic CO lines
expressed in different units(\ceio\ intensities are expressed in units of
main-beam brightness temperature, unlike $\twco$ and $\thco$) provides good
support to the hypothesis made in our analysis.

In the analysis of the dense gas, as traced by CS, we have adopted a lower
value for the kinetic temperature ($20\K$), comparable to the cold dust
temperature derived from the spectral energy distribution of the cores detected
in the survey (see above). The density and column density of the high-density
gas were estimated from a simple analysis of the CS line emission in the
Large-Velocity Gradient approximation. The linewidth was obtained from a
Gaussian fit to the line profiles of the \jtwotoone\ and \jthreetotwo\
transitions. We have assumed a standard abundance $\rm [CS]/[\htwo]= 10^{-9}$.

The dynamics of the molecular cloud gas will be discussed in detail on the
basis of our molecular line emission survey and presented in a forthcoming
paper (Lefloch 2008, in prep.).

\section{Results}

\subsection{Dust Filaments}

\begin{table*}
\begin{flushleft}
\caption[]{Physical properties of the filaments. }
\begin{tabular}{|l|c|c|c|c|c|c|c|c|c|} \hline
Name & Length & Diameter & $\sigma$ & Mass  & N(H) & Density &  Fragment sep.
& Mass per u.l. & Virial mass  \\
     & pc     & pc     & $\kms$ & $\msol$ &  $\cmmd$ & $\cmmt$  & pc &
$\msol\pcmu$ & pc \\ \hline
WF   &  11.8  & 1.0    & 0.7 & 2330 &  1.5(22) & $2.0\times 10^3$ &  3.6
& 197 & 220 \\
EL   &  1.9   & 0.47   & 0.65 & 138 & 2.0(22) & $6.9\times 10^3$ &  0.8
& 74 & 190 \\
SL   & 1.6    & 0.45   & 0.5  & 66   & 9.0(21) & $9.0\times 10^3$ &  0.7
& 39 & 104 \\
TC4f & 1.3    & 0.33   & 0.7 & 237  &5.5(22) & $2.5 \times 10^4$ & 0.41 &
177 & 210 \\
\hline
\end{tabular}
\end{flushleft}
a(b) stands for $a\times 10^b$.
\end{table*}

The presence of "stellar incubators" in M20 was  searched for by mapping the
distribution of the cold dust emission at 1.3mm. We have mapped the emission of
the parental cloud over an area of about 15 arcmin in R.A. by 20 arcmin in
declination. The final map is presented in Fig.~\ref{iramsurvey}. It offers a
very contrasted picture of the nebula with the best angular resolution
available in our millimeter wavelengths observations ($11\arcsec$). All the
condensations discovered in the survey are distributed in a complex network of
filamentary structures. Some of these filaments are detected as dark lanes
against the bright nebular background in the optical. The surface layers of the
UV-heated filaments in the nebula are detected in the mid-infrared with ISOCAM
(CL98) and SPITZER (Rho et al. 2006). Four main filaments could be identified
on the front side of M20 (Fig.~\ref{iramsurvey})~: the lanes which trisect the
nebula (Southern, Eastern, Western lanes) and a long filament on the Western
side ($12\pc$), detected in absorption in the optical, to which we will refer
as the ``Western filament'' (WF; see also Lefloch et al. 2003). Observations of
the molecular gas emission has revealed several other filaments of similar
length, $10\pc$ or more, extending far beyond M20, both in the Northern and
Southern directions, on the rear side of the nebula. The TC4 region was
previously characterized as a compressed shocked layer, filamentary-like, of
accumulated material Southwest of the nebula (LC00), dubbed ``TC4f''. Our
survey shows that the filament stretches far away from M20, to a region of
diffuse emission, South of the nebula.

The physical properties of the filaments were derived from the millimeter
thermal dust emission. The average gas column density $\rm N(H)$ was obtained
from the average flux integrated in the region defined by the intensity contour
$10\mjy/11\arcsec$ beam and is of the order of a few $10^{22}\cmmd$. The
average density was derived  dividing the gas column density by the mean
filament diameter, estimated from the contour $10\mjy/11\arcsec$ beam. To ease
comparison with models, we also computed the mass, mass per unit length of each
filament, and determined the sound speed from the \ceio\ line profiles. The
physical properties of the filaments are summarized in Table~2. The filaments
are rather dense, from a few $10^3\cmmt$ to about $10^4\cmmt$ in the central
regions. WF is noticeably larger and more massive than the  other filaments,
but the molecular gas density and the kinetic energy density are very similar
in all of them. The physical properties of WF, such as its size, its mass per
unit length,  compare very well with thoss of the Orion filaments (Maddalena et
al. 1986) and the infrared dark clouds (IRDCs; see Beuther et al. 2007 for a
review).

All the filaments display evidence of fragmentation at small and large scales.
WF is fragmented in a long chain of massive condensations which stretches down
to the W28 supernova remnant (SNR), located at a distance of 1.9 kpc, similar
to M20 (Vel\'azquez et al. 2002).  We have estimated the mass of each
condensation by integrating on the contour of millimeter continuum emission at
$10\mjy/11\arcsec$~beam; we obtain masses of several hundreds $\msol$ for each
condensation (see Table~4; also Sect.7). Evidence of star formation was
previously found in WF, thanks to the discovery of TC3 (LC00). Several water
masers were discovered in TC3, one of which is associated with a SPITZER
protostellar object (Healy, Hester \& Claussen, 2004).

The detection of Class
0/I protostellar sources in TC1 and TC8 testifies that star formation is going
on in the Eastern lane. The Southern Lane is an exception as it does not
display any hint of star formation activity although two condensations were
discovered in the filament. Such non star-forming (pre-stellar) cores play an
important role as they bring information on the physical conditions of the
parental cloud, before the onset of star formation. Unfortunately, detailed
molecular observations  are available only for a few of them
(TC0,TC6,TC10,TC16).

\subsection{Dust Cores in M20}

\begin{table*}
\begin{flushleft}
\caption[]{Condensations discovered at millimeter wavelengths near M20.
Star-forming condensations are marked by with a $\dag$. }
\begin{tabular}{|l|c|c|c|c|c|c|c|c|c|} \hline
Name & $\alpha(2000)$ & $\delta(2000)$ & Dimensions & Core & Core Virial
Mass$^{2}$ & Fragment & (Virial)$^{1}$ Mass & N(H)$^{2}$
&   \\
     &   &   &   &  ($\msol$) & ($\msol$)& ($\msol$) &($\msol$) &
 ($\cmmd$)  & \\ \hline
TC00 & 18:01:53:63 & -22:50:57.6 & $160\arcsec$ & 49 & 56/93
 & 250 & 213/355  & 7.0(22) & WF \\
TC0  & 18:02:06.35 & -22:57:13.6 & $180\arcsec \times 80\arcsec$ &
\_ & \_ & 474 & 196/327 & 9.5(22) & WF \\
TC0A  & 18:02:06.35 & -22:57:13.6 & $53\arcsec \times 24\arcsec$ &
49 & 58/97 & \_ & \_ & 9.5(22) & WF \\
TC0B & 18:02:01.70 & -22:57:00.0 & $21\arcsec \times 9\arcsec$ &
11& 15/24 & \_ & \_ & 5.7(22) & WF  \\
TC0C & 18:02:03.42 & -22:57:27.6 & $17\arcsec \times 9\arcsec$ &
8 & 16/27 & \_  & \_ & 5.7(22) & WF \\
TC0D & 18:02:04.00 & -22:57:55.5 & $15\arcsec \times 15\arcsec$ &
9 & 19/33 & \_ & \_ & 5.2(22) & WF \\
TC1$\dag$  & 18:02:24.68 & -23:01:17.7 & $46\arcsec \times 46\arcsec$ &
28 & 23/38 & 69 & 55/91 & 1.6(23) &  E \\
TC2$\dag$$^4$ & 18:02:28.49 & -23:03:56.9 & $50\arcsec \times 36\arcsec$ &
27 & 16/27 & 63& 35/59 & 1.6(23) & \_ \\
TC3$\dag$$^5$ & 18:02:05.57 & -23:05:29.0 & $200\arcsec \times 110\arcsec$ &
\_ & \_ & 750 & 294/489 & 3.7(23) & WF \\
TC3A$\dag$$^5$ & 18:02:05.74 & -23:05:32.5 & $20\arcsec \times 20\arcsec$ & 90
& 39/64 & \_ &
\_ & 3.7(23) & WF \\
TC3B$^5$ & 18:02:04.07 & -23:06:06.6 & $13\arcsec \times 13\arcsec$ & 17 &
24/42 & \_ & \_ &
1.3(23) & WF \\
TC3C$\dag$$^5$ & 18:02:05.87 & -23:05:03.9 & $15\arcsec \times 15\arcsec$ &
34 & 30/49 & \_ & \_ & 1.9(23) & WF \\
TC4 & 18:02:12.77 & -23:05:46.7 & $160\arcsec \times 40\arcsec$ &
\_ & \_ &  237 & 159/265 & 2.8(23) &  \_ \\
TC4A$\dag$$^5$ & 18:02:12.77 & -23:05:46.7 & $32\arcsec \times 16\arcsec$ &
58 & 31/51  &  \_ & \_ & 2.8(23) &  \_ \\
TC4B$\dag$$^5$ & 18:02:14.43  & -23:06:47.5 & $12\arcsec \times 12\arcsec$ & 11
& 16/27 & \_ &
\_ & 8.6(22) &   \_ \\
TC4C$^5$ & 18:02:17.00  & -23:07:12.9 & $13.5\arcsec \times 13.5\arcsec$ & 8 &
18/30 & \_ & \_ &
6.5(22) &  \_ \\
TC5$\dag$  & 18:01:57.65 & -23:12:44.6 & $140\arcsec \times 80\arcsec$ &
177 & 134/223 & 807 & 850/1418 & 9.9(23) &  WF \\
TC6  & 18:02:03.37 & -23:08:40.6 & $120\arcsec \times 50\arcsec$ & \_ &
\_ & 43 & 126/210 & 3.6(22) &  WF \\
TC6A & 18:02:02.83 & -23:09:39.2 & $11\arcsec \times 11\arcsec$ &
3 & 18/30 & \_ & \_ & 2.9(22) & WF\\
TC6B & 18:02:01.94 & -23:09:22.0 & $11\arcsec \times 11\arcsec$ &
2.5 & 18/30 & \_ & \_ & 2.5(22) & WF\\
TC6C & 18:02:03.36 & -23:08:39.7 & $11\arcsec \times 11\arcsec$ &
2 & 18/30 & \_ & \_ & 2.3(22) & WF\\
TC6D& 18:02:03.01 & -23:08:25.7 & $11\arcsec \times 11\arcsec$ &
2 & 18/30 & \_ & \_ & 2.0(22) & WF\\
TC7$\dag$ & 18:02:03.50 &-23:09:51.2  & $160\arcsec \times 100\arcsec$ &
17 & 39/65 & 63 & 102/170 & 8.8(22) &  WF \\
TC8$\dag$  & 18:02:30.38 & -23:00:22.2 & $88\arcsec \times 34\arcsec$ &
13 & 14/24 & 46 & 66/109 & 7.0(22) &  E \\
TC9$\dag$  & 18:02:27.25 & -23:03:20.3 & $17\arcsec \times 13\arcsec$ &
\_ & \_ & 4 & 7/12 & 1.6(22) &   \_ \\
TC10 & 18:02:22.3 & -23:02:56.0 & $24\arcsec \times 9\arcsec$ &
2.5 & 5/8 & 25 & 30/51 & 2.3(22) &  \_ \\
TC11 & 18:02:20.4 & -23:01:24.0 & $44\arcsec \times 28\arcsec$ &
\_ & \_ & 6  & 23/38  &  9.0(21)  &  S  \\
TC12$\dag$ & 18:02:31.66 & -22:56:46.2 & $56\arcsec \times 32\arcsec$ &
\_ & \_ & 9 & 76/126  & 9.1(22) &  \_ \\
TC13$\dag$ & 18:02:40.73 & -23:00:30.6 & $55\arcsec \times 35\arcsec$ &
\_ & \_ & 3.5 & 65/108 & 1.2(22) &  E \\
TC14$\dag$ & 18:02:47.26 & -23:06:13.0 & unres. & \_ & \_ & \_ & \_&
4.7(21) &  \_  \\
TC15$\dag$ & 18:02:34.17 & -23:06:53.0 & $21\arcsec \times 11\arcsec$ &
1.4 & \_ & 4.7& 18/30 & 9.3(21) &   \_ \\
TC16$\dag$ & 18:02:34.46 & -23:08:06.4 & $36\arcsec \times 26\arcsec$ &
1.8 & \_ & 9.4 & 24/49 & 1.2(22) &  \_ \\
TC17 & 18:02:16.85 & -23:00:51.8 &  $11\arcsec \times 11\arcsec$ & \_ & \_ & 7
& 12/20 &
4.4(22) & W \\
\hline
\end{tabular}
\end{flushleft}
$^{1}$~The virial mass was computed assuming an isothermal sphere with density
$\rho\propto r^{-2}$~: $M_v= 3 R\sigma^2/G$ and uniform density $M_v= 5
R\sigma^2/G$; the velocity dispersions $\sigma$ is derived from $\ceio$
line observations, or \thco\ if $\ceio$ is not detected.\\
$^{2}$ Peak column density; \\
$^{3}$ WF= Western filament; E= Eastern Lane; W= Western Lane; S= Southern
Lane \\
$^4$ L02. $^5$ LC00\\
\end{table*}

The continuum sources and their physical properties (size, mass, column
density) are listed in Table~3. The sources can be divided into two
categories~: \\
- condensations (or massive fragments) with a typical size of $100\arcsec$
($1\pc$) and a mass of several hundred $\msol$. Their extent is defined by the
contour at $10\mjy/11\arcsec$ beam. \\
- cores with a typical size of a few $10\arcsec$ ($0.1\pc$) and a mass of a few
ten of $\msol$. They are subunits of the fragment and  their extent is defined
by the contour at half power.

All the cores detected in our survey have a typical size in the range
$10\arcsec-20\arcsec$ ($0.1-0.2\pc$). The gas density at the continuum emission
peak ranges between $10^5\cmmt$ and a few $10^6\cmmt$. The mean core density is
higher by a factor of 10, typically, in the regions where evidence of shock has
been detected.

Of the 33 condensations identified in our survey and listed in Table~1, 16
sources are associated with SPITZER protostellar candidates (Rho et al. 2006).
Most of the young stellar objects identified by SPITZER are not associated with
large amount  of cold molecular gas or cold dust emission in our survey ($\rm
N(\htwo) \geq 10^{22}\cmmd$). This is the case of TC14, for which we failed to
detect any associated continuum or line emission and the case of the young
stellar objects IRS~6 and IRS~7 discovered by L01; they exhibit the $9.7\mu m$
silicate band in emission, a direct evidence of the weak amount of parental
material still present. We have searched for outflow signatures towards the
protostar candidates. In most of the cases, confusion from the ambient
molecular cloud is such that it is not possible to discriminate the outflow
wing emission in tracers as common as CO. It is necessary to observe tracers
with a more constrasted emission, hence less abundant, to detect the signature
of the outflowing gas. Unfortunately, it becomes more difficult to detect this
emission in low-density gas region, such as the cometary globules located on
the Eastern side of M20. In a few sources, however, we detected the
high-velocity wings of protostellar outflows  in the CS and SiO line profiles
(TC3,TC4,TC5). We also detected a protostellar Herbig-Haro jet associated with
the protostellar core TC2 (CL98; Rosado et al. 1999;L02).

\begin{table*}
\begin{flushleft}
\caption[]{Physical properties of the cores in the parental
cloud of the Trifid. }
\begin{tabular}{|l|c|c|r|c|c|l|c|c|} \hline
Name &  Emission Peak & Tk &  N($\htwo){}^1$ & $\rm N(\thco)$ & N(CS) & $n(\htwo)^{2}$ &
Mcore$^{1}$ & Location  \\
     &   & (K)   & ($\cmmd$) & ($\cmmd$) & ($\cmmd$) & ($\cmmt$) & ($\msol$) & \\
\hline
TC0A  & ($-238\arcsec, +294\arcsec$)& 25 &  4.9(22)& 5.5(16) & 1.8(13) & 5.5(5) & 49
& WF \\
TC0B  & ($-302\arcsec, +292\arcsec$)& 25 &  2.9(22) & 4.0(16)   &  1.0(13) & 4.0(5) &  11
  & WF \\
TC0C & ($-268\arcsec, +275\arcsec$) & 25 & 2.9(22) & 4.5(16) &  1.1(13) & 4.0(5) &  8
  & WF\\
TC0D & ($-270\arcsec, +252\arcsec$) & 25 & 2.9(22) & 3.8(16)  & 6.0(12)  &
4.0(5) & 9
  & WF\\
 TC1  & ($+16\arcsec, +33\arcsec$)   & 30 & 8.4(22) &  3.6(22) & 1.9(13) &
3.0(5) & 28  &
EL \\
TC2$\dag$ & ($+72\arcsec, -120\arcsec$) & 30 & 8.0(22) & 4.0(16) & 1.8(13) & 3.0(5) & 27
  & South \\
TC3A$\ddag$ & ($-246\arcsec, -220\arcsec$) & 20 & 1.9(23) & 6.0(16) & 2.0(13) & 1.6(6)
 & 90 & WF \\
TC4A$\ddag$ & ($-150\arcsec, -246\arcsec$) & 20 & 1.4(23) & 6.0(16)  & 3.3(13) & 6.0(5)
 & 58 & WF\\
TC5$^4$ & ($-360\arcsec, -640\arcsec$) & 20 & 5.4(23) & \_ & 1.0(14) & 2.3(6)
&  177 & WF \\
TC6  & ($-280\arcsec, -440\arcsec$) & 30 & 1.3(22) & 1.5(16) & 1.2(13) & 8.0(4) & 2-3 &
WF \\
TC8  & ($+94\arcsec, +87\arcsec$) & 30 & 3.5(22) & 5.1(16) & 1.5(13) & 4.0(5) & 6 &
EL \\
TC9  & ($+50\arcsec, -90\arcsec$) & 30 & $\leq 1.0(22)$ & 1.6(16) & \_ & 3.0(4)$^1$ &
4 & South \\
TC10 & ($-20\arcsec, -66\arcsec$) & 25 & 1.2(22) & 2.4(16) & 1.5(13) & 2.5(5) & 25 & SL \\
TC11 & ($-42\arcsec, +30\arcsec$) & 25 & \_ & 1.4(16) &  \_  & 1.0(4)$^1$ & 6 & SL\\
TC13 & ($+240\arcsec,+70\arcsec$) & 25 & 6.0(21) & 1.0(16) & 6.0(12)&1.0(5) & 3.5 & EL \\
\hline
\end{tabular}
\end{flushleft}
$^{1}$ determined from dust\\
$^{2}$ determined from CS \\
${}^4$  Lefloch et al.(2008). \\
$\dag$ L02. $\ddag$ LC00.
\end{table*}

\begin{table*}
\begin{flushleft}
\caption[]{Line parameters of the various transitions observed (intensity in K,
velocity peak and linewidth in $\kms$) at the brightness peak of the
condensations. The intensity of the \ceio\ and CS lines is expressed in units
of main-beam brightness temperature. The symbol ``-'' implies that the relevant
transition has not been observed. NG means that the line profile is too complex
to be fit by simple Gauss functions.
 }
\begin{tabular}{|l|c|c|c|c|c|c|c|c|} \hline
Name &  Emission Peak & CO  &  CO  & CO
& $\ceio$   & $\thco$ & CS  & CS  \\
     &   & $\jonetozero$  & $\jtwotoone$ & $\jthreetotwo$ & $\jonetozero$ &
$\jonetozero$ & $\jtwotoone$ & $\jthreetotwo$\\
\hline
TC0A & ($-238\arcsec, +294\arcsec$)& 22.0 & 23.7 & 12.5 & 1.7 & 12.4 &
2.5 & 3.1\\
      &                            & 10.5 & 10.2 &  10.0 & 10.4 & 10.3 &
 10.5 & 10.5 \\
      &                            & 3.0  & 3.0  &  3.0  & 1.8  &  1.8 &
1.8 & 1.5  \\
      & & & & & & & & \\
TC0B  & ($-302\arcsec, +292\arcsec$)& 22.5 & 22.5 & 16.4  & 1.7 & 14.7 &
2.30 & 2.67   \\
      &                             & 9.9     & 10.0 & 10.2  & 10.6 & 10.3 &
10.60 & 10.55   \\
      &                             & 2.8     & 2.8  &  3.4  & 0.8  & 1.5 &
1.0 & 0.95   \\
      & & & & & & & & \\
TC0C & ($-268\arcsec, +275\arcsec$) & 20  & 22.0  & 13.8 & 1.5  & 11.2 &
2.5 & 2.75\\
        &                          & 10.9 & 10.9 & 11  & 10.6 & 10.5 &
10.5 & 10.5 \\
      &                             & 1.7 & 1.7 & 2.0  & 1.0 & 1.3  &
1.2 & 1.1 \\
      & & & & & & & & \\
TC0D &  ($-270\arcsec, +252\arcsec$)  & 20.0 & 19.0  & 15.7  & 1.1  & 6.8  & 1.2  & 1.4 \\
     &                 & 9.8  &  9.8  & 10.1  & 10.2 & 9.8  & 10.4 & 10.6 \\
     &                 & 3.8  &  4.0  &  3.5  & 1.2  & 1.8  & 1.4  & 0.9 \\
     &                 &  &  &   &  &  &       & \\
TC1  & ($+16\arcsec, +33\arcsec$)   & 19.6 & 25.4  & 15.9  & 1.4 & 12  &
3.0 & 3.0\\
                              &       & 3.2  & 3.2   & 3.55 & 3.7  & 3.8 &
 3.55 & 3.85 \\
     &                              & 2.6  & 3.2   & 3.2 & 1.3  & 1.7 &
1.0 & 1.1\\
      & & & & & & & & \\
TC2$^1$  & ($+72\arcsec, -120\arcsec$) & 31.0 & 29.7 & 18.7 & 2.3 & 13.0 & 3.0
& 4.4
\\
     &                             & 7.7  & 7.5  & 8.2 & 7.7  & 7.65  & 7.7 & 7.7 \\
     &                             & 2.2  & 2.4  & 2.9 & 1.0   & 1.6  & 1.3  & 1.3 \\
     &                             &      &      &     &       &      &      &
     \\
TC3A$^2$ & ($-246\arcsec, -220\arcsec$)& NG     &  NG & NG  &  5.1  & 8.5  &
1.5 & 2.70
\\
     &                             &      &      &     &  21.6 &  21.9 & 22.1 & 20.7\\
     &                             &      &      &     &  1.9  &  2.9 &  5.8$^{\ddag}$ & 1.9 \\
     &                             &      &      &     &      &     &   & \\
TC4A$^2$ & ($-150\arcsec, -246\arcsec$)& NG     & NG  & NG    & 4.3  & 7.4 &
2.1 & 2.5
\\
     &                             &      &      &     & 21.3  & 22.1 & 22.6 & 22.5
     \\
     &                             &      &      &     & 1.2   & 2.3 & 2.7  & 2.6
     \\
     &                             &      &      &     &      &   &      & \\
TC5$^3$  &($-360\arcsec, -640\arcsec$) &  NG    &  NG  & NG & \_ & \_  & 0.34$^+$ & 0.37$^+$\\
     &                             &      &      &     &      &   & 21.1 & 21.4
     \\
     &                             &      &      &     &      &   & 3.0 & 4.6\\
     &                             &      &      &     &      &   &     & \\
 TC6  & ($-280\arcsec, -440\arcsec$) & NG & NG & NG & \_ & 3.1 2.1 &
0.74 0.65$^{\dag}$ & 0.49 0.35$^{\dag}$ \\
     &                               &  &   &  &  &  18.9 16.2 &
20.0 16.5 & 20.0 16.5 \\
     &                               &  &   &  &  &   2.8 2.5 &
3.0 1.6 & 2.8 1.6   \\
      & & & & & & & & \\
TC8  & ($+94\arcsec, +87\arcsec$)   & 21.4 & 21.4  & 15.7 & 1.2 & 9.6  &
1.9  &  2.2    \\
     &                              & 3.8 & 3.6  & 3.85 & 4.1 & 3.7 &
3.5   & 3.8    \\
     &                              & 2.7  & 3.0  & 3.0  & 1.1 & 1.7  &
1.7   & 1.7   \\
      & & & & & & & & \\
TC9   & ($+50\arcsec, -90\arcsec$) & NG  &  NG & NG & 0.40 & 3.5  &  \_ & \_ \\
      &                            &     &     &    & 16.5 & 16.5 & \_ & \_ \\
      &                            &     &     &    & 1.1  & 1.2  & \_ & \_ \\
      & & & & & & & & \\
TC10 & ($-20\arcsec, -66\arcsec$) & 22.8 & 18.9 & 16.5 & 2.0  & 10.0 &
2.5 & \_ \\
     &                            & 0.4  & 0.4  &  0.8 & 0.3  & 0.3  &
0.4  & \_    \\
    &                             & 2.4  & 2.8  & 2.8  & 0.8  & 1.2 &
1.4 & \_ \\
      & & & & & & & & \\
TC11 & ($-42\arcsec, +30\arcsec$) & 16.1 & 25.2 & 16.2 & 1.0 &  2.9
 & 0.62 & \_  \\
     &                            & 1.1  & 1.1  & 1.7  & 1.2 & 1.2
 & 1.1  & \_       \\
     &                            & 2.0  & 2.1  & 2.6  & 0.7  & 1.4
 & 1.4  & \_    \\
      & & & & & & & & \\
TC13 & ($+240\arcsec,+70\arcsec$)   & 20.0 & 23.0 & 16.5 & $< 0.2$ & 4.2 &
1.3 & \_ \\
     &                              & 4.2  & 4.5  & 4.7  &  \_  & 3.8  &
3.8 &  \_ \\
     &                              & 2.3  & 2.9  & 3.0  &  \_  & 1.2  &
1.2 & \_  \\
\hline
\end{tabular}
\end{flushleft}
$^1$ L02. $^2$L00. $^3$ Lefloch et al. 2008 (in prep)
$\dag$~corrected for beam dilution. \\
$^+$ measured on SiO as CS line profiles are self-absorbed.\\
$^{\ddag}$~low resolution spectrum ($1\kms$); contamination from outflow wings
and a second gas component on the line of sight.
\end{table*}

\section{The Western Filament}

Our extended map shows protostellar condensations TC5, TC7 in addition to the
source TC3, previously detected by CL98 and prestellar condensations TC00, TC0,
TC6, distributed in a chain of fragments regularly spaced along WF. The SPITZER
image of the region (Rho et al. 2006) shows that small clusters of a few
protostars have formed in the massive gas fragments  TC3 and TC5. In this
section, we present a detailed analysis of the condensation TC0. The properties
of TC5 and TC6 are discussed in Appendix A.

\begin{figure}
\begin{center}
\leavevmode
\caption{ Millimeter continuum and line emission observed towards TC0
(contours), superposed on a $8\mu m$ image of M20 obtained with SPITZER. In the
lower left  corner of each panel is drawn the beam solid angle (HPFW) of the
observations. For each molecular line, the  emission is integrated between 6.5
and $13.5\kms$.} \label{panel_tc0}
\end{center}
\end{figure}
\smallskip

\begin{figure}
\begin{center}
\leavevmode
\caption{ Thermal dust emission map of TC0 at 1.3mm observed with the IRAM 30m
telescope. First contour and contour interval are 30 and $15\mjy/11\arcsec$
beam respectively.} \label{tc0A}
\end{center}
\end{figure}
\smallskip

Condensation TC0 is located in the Northwest of the Trifid  at the border of
the \Hp\ region (Figs.~\ref{iramsurvey}-\ref{panel_tc0}). In the optical, it is
detected as a region of large obscuration against the diffuse background
emission (see e.g. Rho et al. 2006). This shows that the fragment is located on
the front side of the nebula, which is consistent with the velocity of the main
body gas $v_{lsr}= +10.2\kms$. Dust emission in the mid-infrared shows the
photo-dissociation region as a bright straight bar, with a sharp border at the
interface with the \Hp\ region (Fig.~\ref{panel_tc0}). The molecular gas
tracers confirm the presence of such a strong gradient in the emission.

The cold dust emission at 1.3mm reveals a condensation with typical dimensions
of $100\arcsec \times 80\arcsec$ (Fig.~\ref{tc0A}). The morphology is quite
complex and cannot be easily fit by an ellipse, unlike the other dust
condensations identified in our survey. The contour at half-power delineates
four cores, TC0A-0D (Fig.~\ref{tc0A}). The ``brightest'' one, TC0A, is
noticeably larger than the other condensations, by a factor of 2 typically.
Although the other cores have a comparable brightness, their smaller sizes
result in much smaller masses (typically a factor 4). TC0A-0C-0D are located
right ahead of the ionization front and the photon-dominated region (PDR), as
traced by the $8\mu m$~PAH band. By contrast, the core TC0B is further away
from the PDR and appears to be better protected from the high-energy radiation.

The physical conditions in the  pre-stellar cores TC0A-0B-0C-0D are very
similar, with column densities $\rm N(CS)\sim 1.0\times 10^{13}\cmmd$ and gas
densities $n(\htwo)\simeq 4-5\times 10^5\cmmt$ No evidence for outflow emission
was found either in low- and high-density gas tracers, like in TC3-TC4, which
is consistent with the absence of protostellar activity (see
Fig.~\ref{panel_line_TC0}). Line profiles towards 0B, 0C, and 0D are doubly
peaked. In addition to the main body gas emission, two velocity components are
detected, between $+7\kms$ and $+9.2\kms$, and between $+11\kms$ and
$+12.5\kms$. In particular the kinematic component at "blueshifted" velocities
is detected in the CS line profiles, either as well separated component or as
shoulder to the main body line profile. As can be seen in
Fig.~\ref{velmap_TC0}, both secondary components follow closely the border of
the PDR excited by M20, and most likely trace the signature of the  shock
driven by the ionization front at the surface of the fragment. The widespread
CS emission detected all over TC0 shows evidence for high density gas ($\rm
n(\htwo) \approx 10^5\cmmt$.

\begin{figure}
\begin{center}
\leavevmode
\caption{ Molecular line emission observed towards the cores 0A-0B-0C-0D of
condensation TC0 at offset position ($-238\arcsec$,$+294\arcsec$),
($-302\arcsec$,$+292\arcsec$), ($-268\arcsec$,$+264\arcsec$) and
($-270\arcsec$,$+252\arcsec$), respectively. Fluxes are expressed in units of
antenna temperature. The dashed line marks the velocity of the main body gas.
}\label{panel_line_TC0}
\end{center}
\end{figure}
\smallskip

\begin{figure*}
\begin{center}
\leavevmode
\caption{Emission (contours) of the CO \jtwotoone\ line integrated in the
velocity intervals~: [$7.0;8.8\kms$] (left), [$9.1;11.1\kms$] (middle),
[$11.5;12.7\kms$] (right). The emission is superposed on the emission of the
main body gas (greyscale), integrated in the interval [$9.1;11.1\kms$].
Intensity contours are respectively~: 8,11,14,..$29\K\kms$ (left);
20,28,36,40,$44\K\kms$ (middle), 4,5,7,9,$11\K\kms$ (right).}\label{velmap_TC0}
\end{center}
\end{figure*}
\smallskip

\section{The Eastern Lane}

Three main condensations were identified in the Eastern lane~: TC1, located
close to the central stellar cluster (Fig.~\ref{spitzer_TC1}), TC8 and TC13,
which both  lay further East in the lane (Fig.~\ref{iram_TC1}). A fourth,
starless, condensation was discovered in the filament, peaking at $(-50\arcsec,
-10\arcsec)$. We do not report on its properties in detail as it remains
undetected in our millimeter continuum survey; we mention it for completeness.
TC8 and TC13 are discussed in Appendix B. We discuss in greater detail the
properties of TC1, which is the only cometary globule of the sample for which
mid-infrared spectroscopy could be obtained.

\begin{figure}
  \begin{center}
    \leavevmode
\caption{ Cometary globule TC1 as seen in the mid-IR with ISOCAM ($8-15\mu m$)
and SPITZER ($3.6\mu m$, $4.5\mu m$, $5.8\mu m$, $8.5\mu m$). In the bottom
right panel the 1.3mm thermal dust emission (black contours) superposed on a
greyscale image of the $13.3-16.2\mu m$, as observed with ISOCAM using the
Circular Variable Filter. Contours of cold dust emission are 10,20,40,60 to 160
mJy/$11\arcsec$~beam by step of 20  mJy/$11\arcsec$~beam.Contours of mid-IR
emission range from 0.25 to 0.65 mJy/pixel by step of 0.1 mJy/pixel. The arrow
points to the position of the protostellar source. }\label{spitzer_TC1}
  \end{center}
\end{figure}

TC~1 appears as a bright condensation  in the Eastern lane, 40\arcsec\ East of
HD~164492~A (see Figs.~\ref{spitzer_TC1}-\ref{iram_TC1}), both in the
millimeter continuum and in the molecular gas emission. The condensation is
well separated from the diffuse dust lane emission. The contour at half power
defines a central dense core of  diameter of $0.16\pc$ and typical density
$n(\htwo)= 1.8\times 10^5\cmmt$. The outer region, with a diameter of $0.4\pc$
is characterized  by lower densities ($n(\htwo)= 2.5\times 10^4\cmmt$). The
physical parameters of TC1 (gas column density, mass, size, density) are very
similar to those derived for the protostellar condensation TC2 in the South of
the Trifid (see Table~4 and L02).

\begin{figure}
\begin{center}
\leavevmode
\caption{ Millimeter line emission observed in the Eastern Lane towards TC1 and
TC8. The emission is integrated between 0 and $+7\kms$. In the lower left
corner of each panel is drawn the beam solid angle (HPFW) of the observations.
The line emission integrated between 0 and $7\kms$ is shown in the other
panels. For the \thco\ and the \twco\ transitions the first contour and contour
interval are 0.1 times the flux peak value. For the CS \jtwotoone\ line, first
contour and contour interval are $0.5\K\kms$. For the \ceio\ emission, the
first contour and contour interval are 0.2 times the flux peak value. The
positions observed at IRAM are indicated (open squares) in the upper left
panel.} \label{iram_TC1}
\end{center}
\end{figure}
\smallskip

\begin{figure}
  \begin{center}
    \leavevmode
\caption{ Spatial distribution of the UIBs at 6.2, 7.7, $\rm 11.3\mu m$, and
the [NeII] line in the TC~1 region (contours) superposed on an optical $\rm
H\alpha$ image of the region (HST). First contour and contour interval are 0.04
Jy/px and 0.03 Jy/px respectively. }
\label{pah_tc1}
\end{center}
\end{figure}
\smallskip

The emission of the optical bright rim on the Southern border of TC~1 in the
optical image of the Trifid (Fig.~\ref{spitzer_TC1}), as well as the detection
of the [NeII] line over the globule (Fig.~\ref{pah_tc1}) testifies to the
condensation being photoionized. The dark appearance of the lanes implies that
they lie on the front side of the ionized nebula, and hence they are
illuminated on the rear side. The CVF imaging of the UIBs traces the outer
parts of the condensation, which looks like an almost continuous ring
(Fig.~\ref{pah_tc1}). The low emissivity of the central regions can be
attributed to absorption by the cold core detected at millimeter wavelengths
where large dust and gas column densities are measured, corresponding to $\rm
A_v\sim 145$, and $\tau_{12.7\mu m}= 2$, adopting the reddening law derived by
Lynds et al. (1985). We have searched for a mid-IR signature of the protostar
discovered in TC1 by integrating the ISOCAM flux between 13.3 and $16.2\mu m$,
where the radiated flux is the highest in the CVF band, and the extinction,
longwards of the silicate absorption range, is the weakest. The images of the
globule in the IRAC and ISO LW10 filter are displayed in
Fig.~\ref{spitzer_TC1}. A point source is detected very close (within
$4\arcsec$) to the cold dust peak (see arrow in Fig.~\ref{spitzer_TC1}). The
high-sensitivity SPITZER/IRAC images have detected emission from the source at
shorter wavelengths. It is unresolved in these observations
($1\arcsec$-$2\arcsec$). A SED was built using the  IRAC colors and the
MIPS~$24\mu m$ data (see Rho et al. 2006), which yields an estimate of
$570\lsol$ for the source luminosity. This value is typical of an
intermediate-mass protostar.

A spectrum of the source between 5 and $16\mu m$ was obtained with ISOCAM/CVF
after subtracting the emission of the condensation averaged over the closest
neighbouring pixels.  Instrumental problems have affected the long wavelength
border of the spectrum, resulting in a spurious decrease of the flux  beyond
$16\mu m$. The spectrum in the range $5.0-15.5\mu m$ is shown in
Fig.~\ref{cvf_tc1}. The spectrum  exhibits a deep absorption due to the
silicate band at $9.7\mu m$ and a rising slope longward of $10\mu m$,
characteristic of embedded sources. Longward of $15\mu m$, the spectrum
displays another dip, due to absorption by the $\rm CO_2$ ice band, centered at
$15.2\mu m$. The $7.7\mu m$ emission band detected in the TC1 spectrum is
probably due to PAHs excited on the rear side of the condensation, as was
reported in TC2 (L02). The spectrum could be approximately fit by two
black-bodies at temperatures of $250\K$ and $140\K$ modified by dust opacity
law $\tau_{\nu}\propto \nu^{1.3}$, a spectral index value typical in the mid-IR
regime, surrounded by an absorbing cold layer of column density $\rm N(H)=
8.0\times 10^{22}\cmmd$, and a total $\rm CO_2$ ice column density $\rm
N(CO_2)= 3.3\times 10^{18}\cmmd$, using integrated absorption cross sections
measured in the laboratory (see Cernicharo et al. 2000).

In the absence of observations resolving the emission region, we have assumed
the following sizes~:  $0.2\arcsec$ (350 AU)
for the $250\K$ component, comparable to the diameter of the disks detected
in the central stellar cluster (see L01), $0.5\arcsec$ (800 AU) for the lower
temperature component ($140\K$). Under such hypothesis, we estimate
gas column densities $\rm N(H)= 4.0\times 10^{20}\cmmd$ and
$\rm N(H)= 3.7\times 10^{21}\cmmd$, respectively. The spectrum is thus
consistent with the picture of a protostar forming at the center of the
TC1 core.

\begin{figure}
  \begin{center}
    \leavevmode
\caption{ Spectrum in the range $\rm 5-17\mu m$ of the protostar embedded in
TC~1 (thick), after subtracting the ambient gas component. The spectrum is fit
by two black-bodies modified by a dust opacity law $\tau_{\nu}\propto
\nu^{1.3}$at $250\K$ (dashed) and $140\K$ (dotted) respectively. The total fit
is displayed in thick. }
\label{cvf_tc1}
\end{center}
\end{figure}
\smallskip

Some residual emission arises in the UIBs at 6.2 and $7.7\mu m$. This is
reminiscent of the situation in the core of star forming globule TC~2, in which
no emission at all is detected between 5 and $\rm 17\mu m$ apart from a weak
emission peak at the position of the $\rm 7.7\mu m$ UIB. A simple modelling
showed that this emission arises from the PDR on the rear side of the globule,
where photoionization takes place. It is strongly absorbed by the material of
the globule, except near 6.2 and $\rm 7.7\mu m$ where local minima in the dust
absorption curve allow some mid-IR photons to escape. Hence the ring
configuration arises naturally  if, just like TC2, the rear side of TC1
undergoes heavy photoionization, once the surrounding lower-density material
has been photoevaporated. From the absorption produced by the cold dust in the
PAH $6.2\mu m$ band and the [NeII] line, we derive $\tau_{12.7\mu m}\simeq 1.7$
and $\tau_{6\mu m}
> 2.5$. Those values are consistent with the extinction derived from the
millimeter continuum observations and an opacity law  $\tau_{\nu}\propto
\nu^{1.3}$.

A secondary component, detected at $v_{lsr}= +1.8\kms$ could trace the
kinematic signature of the shock front driven by the photoionization of the
surface layers.

\begin{figure}
\begin{center}
\leavevmode
\caption{ Spectra obtained towards TC1 and TC8 in the transitions $\ceio$
$\jonetozero$, $\thco$  $\jonetozero$, $\twco$ $\jonetozero$, $\jtwotoone$ and
$\jthreetotwo$ (from bottom to top). Flux is expressed in units of antenna
temperature. The dashed line marks the velocity of the main body gas.}
\label{panel_line_tc1}
\end{center}
\end{figure}
\smallskip

\begin{figure}
  \begin{center}
    \leavevmode
\caption{ Dense gas emission observed in the lines of CS $\jtwotoone$ and
$\hcop\jonetozero$ at the emission peak of TC1 and TC8, in the Eastern dust
lane. The SEST spectra of TC8 are corrected for the main beam dilution effect.
The IRAM spectra of TC1 are expressed in units of main-beam brightness
temperature.} \label{dense_tc1}
  \end{center}
\end{figure}
\smallskip

\section{The Southern region}

Cometary globule TC2 is the first (proto)stellar incubator identified in the
Trifid thanks to the splendid photoionized jet HH~399 which propagates out of
the head of the globule (see Fig.~\ref{panel_tc2}; also Cernicharo et al.
1998). The dust and molecular line emission of cometary globule TC2, as
observed in the mid-infrared with ISO and at millimeter wavelengths, was
discussed in detail by L02. The signature of the radiatively-driven implosion
of the globule was unambiguously detected in the high-density gas tracers
emission (CS, $\hcop$); numerical modelling of the globule yielded a duration
of $0.3\Myr$ for the photoionization. It comes out that the lifetime of the
globule exposed to photoionization is about $1.9\Myr$, long enough to allow the
protostar to complete the accretion phase.

The ISO data were not sensitive enough to detect emission from the protostar.
Based on the SED obtained with ISO in the range $50-200\mu m$, L02 estimated a
source luminosity $L\sim 500\lsol$, from which they concluded the protostar was
an intermediate-mass candidate. Rho et al (2006) obtained a similar estimate
($590\lsol$) using an $3.6-25 \mu m$ SED constructed from the IRAC and MIPS
colors. The protostar powering the jet was only tentatively identified, based
on VLA observations of the free-free continuum. The much higher sensitivity of
the IRAC images has allowed to detect a few sources with colors typical of
young stellar objects, in the region around cometary globule TC2 (see Rho et
al. 2006). SPITZER detected only {\em one} source with protostellar colors
inside the globule. Interestingly, this source is aligned in the HH~399 jet
direction, and lies close to the cold dust emission peak
(Fig.~\ref{panel_tc2}). It coincides with one of the sources marginally
detected at the VLA (L02). The source detected with SPITZER/IRAC is most likely
the protostar powering HH~399.

\begin{figure}
  \begin{center}
    \leavevmode
\caption{The TC2 region seen in the mid-IR with ISOCAM ($8-15\mu m$) and
SPITZER ($3.6\mu m$, $4.5\mu m$, $5.8\mu m$, $8.0\mu m$). In the bottom right
panel the 1.3mm thermal dust emission (black contours) superposed on a
greyscale image of the ionized gas emission as seen in the optical with HST.
Contours of cold dust emission are 10, 20, 40,60 ...160 mJy/$11\arcsec$. The
arrow points to the position of the protostellar source. }\label{panel_tc2}
  \end{center}
\end{figure}
\smallskip

\begin{figure}
  \begin{center}
    \leavevmode
\caption{ Spectrum in the range $\rm 5-17\mu m$ of TC9, after subtracting the
emission of a reference position. The spectrum is fitted by two modified
blackbodies with a dust opacity $\tau_{\nu} \propto \nu^{1.3}$, at temperatures
of 5000~K (dashed) and 320~K (dotted) respectively. The sum of the two
components is displayed in a thick line. }\label{cvf_tc9}
\end{center}
\end{figure}
\smallskip

ISOCAM detected emission between 8 and $15\mu m$ from a small condensation
($13\arcsec \times 17\arcsec$) in the Northwest of TC2 at offset position
(Fig.~\ref{panel_tc2}).  The condensation is not centered on the flux peak,
which indicates the presence of a second, fainter, source.  The SPITZER/IRAC
images at shorter wavelength indeed unveil two sources located at
$(50.7\arcsec,-89.2\arcsec)$ and $(49.6\arcsec,-97.4\arcsec)$, respectively,
lying in a common envelope of $8\arcsec \times 17\arcsec$.  The two sources
have very different IRAC colors. The Northern source is identified as a Class
I, with IRAC fluxes  similar to those measured towards TC4A, hence compatible
with the birth of an intermediate-mass object. The Southern source has typical
stellar colors  (Rho et al. 2006). It could be a bright field star, located
behind the small dust core, or a young star which has emerged from the
photoevaporated layers, like the EGGs of the Elephant trunks in M16 (Hester et
al. 1996).

Mid-infrared spectroscopy in the range $5-17\mu m$ was obtained with the CVF.
The pixel size was $6\arcsec$. We present here a spectrum of the emission
averaged over the condensation (Fig.~\ref{cvf_tc9}), after subtracting the
emission of a reference position, chosen 1 arcmin North of TC9. The spectrum is
dominated by continuum emission decreasing between 5 and $16.5\mu m$, and a
strong absorption in the $9.7\mu m$ silicate band. This shows evidence for a
high temperature region inside the object. The overall emission can be roughly
reproduced by a black-body law at a temperature of $250-400\K$ modified by a
dust opacity law $\tau_{\nu} \propto \nu^{1.3}$, absorbed by a cold dust layer
with an extinction $\rm A_v= 30$, which corresponds to a gas column density
$\rm N(\htwo)= 1.6\times 10^{22}\cmmd$, adopting the reddening law derived by
Lynds et al. (1985) in the Trifid (Fig.~\ref{cvf_tc9}). The uncertainties in
such a fit are rather large due to the narrow wavelength range observed.  We
found that introducing an additional component tracing photospheric emission at
about 5000~K provides a better fit to the spectrum. Assuming sizes of 1~AU for
the photospheric emission region and of 100~AU for the warm dust component, we
estimate a typical gas column density of $3.2\times 10^{20}\cmmd$ in the warm
dust region. The assumed sizes are arbitrary, as the only constraint we have is
that they are less than $1\arcsec$, the size of the IRAC PSF at $3.6\mu m$. The
ISOCAM spectrum is typical of young stellar objects in the Class II phase and
consistent with the emission arising from a young stellar object surrounded by
a protoplanetary disk, embedded in a small gas condensation. The evolutionary
age of Class II objects is typically several $10^5\yr$ up to a few $\Myr$, i.e.
it compares well with the age of the \Hp\ region. It suggests that such sources
could have formed in the same burst of star formation as the exciting star of
M20. From the gas column density of absorbing material, we estimate a mass of
$4-8\msol$ for the TC9 core.

Searching for molecular gas emission from TC9, we have detected a clump of gas
in the millimeter lines of CO and its isotopologues (see Fig.~\ref{iram_TC9}).
The lines peak at $16.5\kms$ (Fig.~\ref{panel_line_TC9}), which implies that
TC9 is located on the {\em rear} side of the nebula. The clump is hardly
resolved at 1.3mm in the \twco\ \jtwotoone\ transition.  The kinetic
temperature of the molecular gas is difficult to estimate as the condensation
is unresolved by the telescope beam. The $\twco$ $\jonetozero$ line  consists
of a narrow line of 7~K superposed on a broad pedestal of 4~K. Taking into
account the beam dilution effect on the core, we find that the  $\twco$
$\jonetozero$ line brightness has to be less than $35\K$.

\begin{figure}
\begin{center}
\leavevmode
\caption{ Mid-infrared (ISOCAM) and millimeter line emission towards TC9. The
molecular line flux is integrated between 15.25 and $17.25\kms$. Solid circles
indicate the HPBW beam size.  (upper left)~: The mid-infrared $8-15\mu m$ is
shown in black contours. Levels range from 160 to 180 by 10, 200 to 240 by 20,
240 to 360 by 40 MJy/sr. In thin contours is drawn the 1.3mm emission; contours
range from 10 to 30 by 10 and from 40 to 150 by 15 mJy/11\arcsec beam. We
indicate for each panel, the observed transition, the first contour and the
contour interval, respectively. (upper right)~: 0.2 and $0.1\K\kms$. (middle
left)~: 0.4 and $0.2\K\kms$. (middle right)~: 1.6 and $0.2\K\kms$.(bottom
left)~: 15 and $1\K\kms$. (bottom right)~: 14.5 and
$0.5\K\kms$.}\label{iram_TC9}
\end{center}
\end{figure}
\smallskip

\begin{figure}
\begin{center}
\leavevmode
\caption{ Molecular line emission observed towards TC9. Fluxes are expressed in
units of antenna temperature. The dashed line draws the velocity of the main
body gas of TC9. }\label{panel_line_TC9}
\end{center}
\end{figure}
\smallskip

\section{Discussion}

\subsection{Core Properties}

In many cores, self-gravity is comparable to the kinetic energy
available, hence plays an important dynamical role in the evolution of the
cores.
This is the case of cometary globules TC1, TC2, TC8, embedded cores like
TC3A, TC4A, TC5, all of them are star-forming condensations. However,  the non
star-forming condensations TC0A, TC00 are also found in this class. This
suggests they might be on the verge of gravitational collapse.

In most of the pre-stellar condensations (e.g. TC0B,0C,4C,6A-D,7), self-gravity
is much lower than the available kinetic energy. This also applies to the
evolved cometary globules TC12,13,14,15,16 which have already formed
protostars, detected with SPITZER. In such objects, the dense gas emission is
restricted to the region right behind the bright rim and we do not find
evidence of a dense protostellar envelope which would be associated with the
SPITZER sources, like in e.g. the young Class0/I  TC2-3.  Most likely, the bulk
of accretion is over in these cometary globules, as only low column densities
of material are still available there.

Overall, the star-forming cores detected in our survey have masses of $10\msol$
or more (up to $100-200\msol$; see Table~1). The two cometary globules which
have been studied in detail, TC1 and TC2, exhibit very similar physical
properties, in terms of core mass, density, temperature and turbulence. With an
average core density of $1.8\times 10^5\cmmt$ and sound speed $\sigma=
0.42\kms$, the Jeans length is $\lambda_J$ is $\approx 0.17\pc$ in TC1 and TC2,
hence approximately the core diameter.  It is therefore no wonder that only one
protostar was
detected in the condensations. 
In other cases, the Jeans length  $\lambda_J$ is less than one core radius  and
the Jeans mass is a small fraction of the total mass available in the
condensation. Hence, the cores appear to be dense and massive enough to give
birth to  a few stellar objects.  This is in agreement with the IRAC maps of
TC3, TC4, and TC5 which unveil clusters of protostars.

On the contrary, most of the non star-forming (pre-stellar) condensations have
masses of a few $\msol$, and self-gravity is not  large enough to
counterbalance the kinetic energy of the gas. The only exceptions are  TC0A and
TC00 ($49\msol$).  We conclude that the cores in TC0 and TC00 are privileged
sites for the formation of the next generation of stars. We speculate that the
gravitational collapse will start as soon as the cores will be compressed by
the shock preceding the ionization front. The  prestellar cores in TC0 and
exhibit physical properties (mass, temperature, density) similar to those of
the star-forming cores TC1 and TC2. Hence, we expect that most of the new
forming stars will be low- (and/or intermediate) mass stars and will form
inside these fragments.

\subsection{Evolutionary Sequence for the cores}

The cold dust cores detected in our survey are submitted to  a large variety of
physical conditions depending on the photoionization  in the parent cloud~:
very deeply embedded cores in massive molecular gas fragments (e.g.  TC3-4-5);
cometary globules, i.e. cores  surrounded by a dense envelope exposed to the
ionizing nebular radiation (e.g. TC1-2); cores in an advanced stage of
photoionization, which led them to get rid of their envelope (e.g. TC12-13).

Undoubtedly, such sources are in a very advanced stage of protostellar
evolution and photoionization. Some other sources, such as TC9 and TC13, are
associated with molecular material detected in line emission, but remain below
the detection limit of our millimeter continuum survey.

Our survey allows us to draw the full evolutionary sequence of cometary
globules exposed to the ionizing nebular radiation, from their initial stage of
formation, as illustrated by the pre-stellar cores embedded in the fragments of
dense molecular gas TC0 and TC00, to the late stages when the bulk of the
envelope has been photoevaporated, like in TC9. Condensations TC1-2-8, directly
exposed to the harsch nebular ionizing radiation, are still emerging from the
low-density layers of the parental molecular cloud. This differs from the
situation in M16, another well-studied \Hp\ region, where the erosion of
molecular gas condensations by the ionizing radiation has shaped ``pillows'' in
the parent cloud.

The free-free emission was observed towards TC1 at the VLA by Yusef-Zadeh et
al. (2005), who derived an electron density of $\approx 2000\cmmt$ in the
bright rim, i.e. similar to the value measured in the bright rim of TC2. Taking
into account that the other properties are similar, the mass-loss rate induced
by photoevaporation should be close to the value derived at the surface of
TC2~: $34\msol\Mymu$. The lifetime of TC1 is therefore $\approx 2\Myr$, similar
to the expected lifetime of an O~7.5 III star (Martins et al. 2005). The
photoevaporation timescale of the condensations TC1 and TC2 is long enough to
allow the central protostellar objects to accrete the bulk of material from
their envelope before being exposed to the ionizing radiation. Hence, the
incubators located in the densest part of the molecular cloud, in the dust
lanes of the Trifid, will eventually allow their young stellar objects to reach
``maturity'' (the protoplanetary phase) before being exposed to the harsh
ionizing radiation.

The fate of TC9, which is located in a less dense region on the rear face of
the nebula, is somewhat different. TC9 consists of  a small core of dense gas
and dust. The central source is surrounded by sufficiently large amount of
materials that the $9.7\mu m$ silicate band is still detected in absorption.
Besides, the SED of the central source is typical of Class~II objects,
consistent with the active accretion phase being now terminated. We adopt as a
rough estimate of the ionizing flux impinging on TC9 the one measured at the
surface of TC2~: $\Phi\sim  10^{10}\cmmd\smu$. This is justified because TC2
and TC9 have similar projected distances to the exciting star of the nebula.
Using Eq. (36) of Lefloch \& Lazareff (1994), which relates the mass-loss rate
to the size of the condensation  and the ionizing photon flux, we find a mass
loss rate of $\simeq 8\msol\yrmu$, where we have adopted a mean radius of
$\simeq 8\arcsec$; hence, the condensation will have been photoevaporated in a
few $10^5\yr$, while the young stellar objects TC1 and TC2 will still be
shielded from the ionizing radiation. TC9 is a clear case of a protostellar
core overrun by the ionization front, which is about to disappear in the
nebular gas.

IRS~6-7, TC11-12,13 illustrate the ultimate stages of early stellar evolution
when the parental envelope of the newly born stars has been almost fully
photoevaporated and hardly any cold molecular gas emission is detected. In
these objects, self-gravity fails to bind the structure, as evidenced
by the ratio between the virial and the actual mass of the condensations.
We note however that the gravitational potential due to the central object(s)
was ignored, so that the actual discrepancy might be less.

\subsection{Core formation and Filaments}

The overall distribution of molecular material differs a lot from the free-free
emission distribution observed with the VLA, which traces an almost perfect
Stromgren sphere (CL98). Most of the protostellar cores are encountered in
filamentary structures. Some of them, like TC1, TC2, TC4, are detected in the
large-scale photon-dominated region but its not the case for all of them. In
particular, cores and fragments in WF, such as TC00, 6, 7 are encountered far
away from the \Hp\ region.

Since the WF filament stretches over $\approx 10\pc$, far beyond the ionization
radius of M20 and connects several complexes together, we can exclude that it
formed from the fragmentation of the layer of molecular material swept-up in
the expansion of the \Hp\ region, as in the ``collect and collapse'' scenario
(Elmegreen \& Lada 1977; Elmegreen 1989). WF was already present before the
ignition of the \Hp\ region, and so were probably the dust lanes of the Trifid,
and other filaments, like in the TC4 region. The separation of the massive
condensations in WF along the filament is approximately constant, equal to
$1.1\times 10^{19}\cmmd$ ($3.6\pc$); this provides additional evidence for the
fragmentation scenario.

In the filaments EL, SL, and TC4f, the size of the cores is very close to the
diameter of these filaments, and their Jeans length. This fact supports the
idea that the cores form as a consequence of the fragmentation of the
filaments. Also, many of the pre-stellar cores, such as TC6,10,11 are not
gravitationallybound, which implies that they probably formed dynamically
inside the filaments.

We explore here the scenario that all the condensations detected in our survey
actually result from the fragmentation of  filamentary clouds. These questions
will be addressed more thoroughly in a detailed analysis of the molecular gas
dynamics in a forthcoming article (Lefloch 2008, in prep).

\subsection{Fragmentation of WF}

The map of the thermal dust emission shows that the material of WF is
concentrated mainly in the massive condensations TC00, TC0, TC3, TC5. From the
average gas velocity dispersion in the TC0-TC3-TC6 region ($\sigma= 0.7\kms$),
we estimate the virial mass per unit length $\rm m_v= 2 \sigma^2/G =
220\msol\pcmu$, which compares well with the mass per unit length in WF
(Sect.~4). We conclude that the filament was initially self-gravitating. Under
the simplifying hypothesis that WF was initially in hydrodynamical equilibrium,
and that it fragmented under some external perturbations, the properties of the
fragments will be mainly determined by the fastest growing mode of the
perturbation.

We present here a simple analysis,  based on the results of the continuum
survey. The characteristic size for a filament in hydrodynamical equilibrium is
given by~:
\begin{eqnarray*}
H & = & 0.0035 \left(\frac{c_s}{0.3\kms}\right)
\left(\frac{n_0}{2\times 10^6\cmmt}\right)^{-1/2} \pc\\
\end{eqnarray*}
where $c_s$ is the effective sound speed (= 1d velocity dispersion $\sigma$)
and $n_0$ is the gas density in the central region of the filament (see
Nakamura, Hanawa \& Nakano, 1993), and is approximately twice the average
density of the filament. Hence, we  find $\rm H= 0.18\pc$ and an effective
filament radius of $\rm R_e= 2\sqrt{2}H = 0.51\pc$. The actual diameter of the
filament is not so easy to measure from the emission maps; the initial diameter
of the filament could have been larger before undergoing external compression.
The observations at SEST and CSO show a somewhat larger  radius  (of about
$0.6-0.8\pc$), though not inconsistent with these results. For a purely
hydrodynamically supported filament, the most unstable mode of axisymmetric
perturbations has a wavelength $\lambda= 22.0 H \simeq 4.0\pc$, a value  in
reasonable agreement with the mean separation between fragments $\approx
400\arcsec= 3.6\pc$.

A better agreement is obtained under the assumption that the filament is
threaded by a magnetic field  and that equipartition is reached between kinetic
and magnetic energies. The role of the magnetic field must be taken into
account as it provides additional support against the filament self-gravity; as
a consequence the filament increases its  scale height and its effective radius
$\rm R_e$. With the same physical parameters and assuming equipartition between
magnetic and kinetic energy,  the scale height $H\prime$ becomes $\rm
H^{\prime}= H \times \sqrt{2}= 0.26\pc$ and the equivalent radius $R_e=
0.73\pc$ ($90\arcsec$). The wavelength of the most unstable mode  for
axisymmetric perturbations in a magnetized filament is then $\lambda = 13.0
H^{\prime}= 3.4\pc$. Overall, these results agree well with the observations,
in particular the constraint on the diameter of the filament is better
fulfilled.

Inspection of the fragment along WF shows a marked trend. Whereas TC3 and TC5
fragments have very comparable masses ($\approx 800\msol$), the mass of each
newly encountered fragment  when moving North along the filament decreases, by
a factor about 2 (see Table~2). The material distribution of TC00 seems less
structured than in TC0, itself less structured than in TC3 and TC5. This
suggests that the fragmentation of the filament could have started first with
the formation of TC3 and TC5, then proceeded to the North of the filament,
where TC0 and TC00 would represent subsequent generations of fragments.

The sequential star formation triggered by the expansion of an \Hp\ region,
impacting a filamentary cloud has been studied numerically by
Fukuda \& Hanawa (2000).
Their simulations show that under a wide range of physical conditions  it
takes about 10 times the radial sound crossing time
\begin{eqnarray*}
\tau & = &  H/c_s \\
& = & 1.7\times 10^4 \left(\frac{n_0}{2\times
10^6\cmmt}\right)^{-1/2}\yr
\end{eqnarray*}
to form the first condensations in the shocked filament. In the case of WF,
this time would be as large as $4\Myr$, almost one order of magnitude larger
than the age estimated for the Trifid (L02). Despite all the uncertainties in
the age determinations, the detection of photoionized protoplanetary disks in
the central cluster of the \Hp\ region (L01) implies that the nebula is young.
It is more  likely that fragmentation in WF was initiated by some other events
than the interaction with M20. An interesting result of these simulations  is
that fragmentation can occur at much earlier times as more kinetic energy is
injected in the shock. An alternate scenario is that fragmentation was already
ongoing before the ignition of the nebula.

A large number of observations support the existence of a physical interaction
between the young W28 SNR and the surrounding molecular gas (Wootten 1981,
Frail \& Mitchell 1998, Arikawa et al. 1999; Rho \& Borkowski 2002; Yusef-Zadeh
et al. 2000; Reach, Rho \& Jarrett). It is not clear however if such events as
energetic as supernova remnants could trigger the fragmentation of the filament
on such short timescales. On the other hand, the characteristic timescales of
TC3 and TC5 are fully compatible with  a dynamical interaction between WF and
W28. As discussed above, the free-fall gravitational timescale in the cores of
TC3 and TC5 are extremely short, of the order $10^4\yr$, and similar to the age
of the protostellar sources in TC3 and TC5. This is in excellent agreement with
the estimated age of the W28 supernova remnant ($3.3\times 10^4\yr$,
Vel\'azquez et al. 2002).

In order to confirm this hypothesis, it should be investigated whether the
impact of the SNR shock on  the WF filament, the shock could trigger the
gravitational collapse of the core and the formation of the protostellar
cluster.  The location of TC5 at the edge of the S28 shell star formation
strongly suggests that star formation has indeed been triggered in the core as
soon as the supernova shock hit the condensation.

\begin{figure}
  \begin{center}
    \leavevmode
\caption{ Map of the 1300$\mu m$ thermal dust emission superposed on a map of
the  20cm free-free emission of the region in greyscale (from Yusef-Zadeh et
al. 2000).}\label{resumen}
\end{center}
\end{figure}
\smallskip

\subsection{Fragmentation of TC4f}

In TC4f filament, the different cores (4a,4b,4c,4d) are located along the
filament, at the border of the nebula. A detailed analysis of the physical
conditions showed evidence for compression of the filament gas (LC00). No
evidence of velocity gradient was found between the various molecular gas
tracers.  If the collapse of the cores had started before
 compression of the surface layers occurred, lines with different optical
depths should probe different regions of the core, hence  tracing regions at
different velocity. There is no such velocity gradient observed, between the
optically thick CS  and the optically thin \ceio\ line. Most likely, the cores
TC4a-b-c-d formed after compression of the filament. The properties of the
cores  are consistent with them being fragments of this layer (LC00). The
timescale for gravitational collapse of a decelerating shocked gas layer is
$\tau \sim 0.25(G\rho_0)^{-1/2}$, where $\rho_0$ is the typical (pre-shock)
density of the parent molecular cloud (Elmegreen, 1989). The average density
determined over WF ($2000\cmmt$) provides a good estimate of $\rho_0$. The
typical timescale for the fragmentation of TC4f density is therefore
$0.35\Myr$, compatible with the age estimated for M20. To summarize, there is
good observational evidence that TC4f is a filament compressed by the expansion
of the \Hp\ region, which subsequently became gravitationally unstable and
formed young stellar objects.

\subsection{Fragmentation in the dust lanes}

We have carried out a similar analysis for filaments SL and EL. The results are
summarized in Table~3. Unlike WF and TC4f, the Southern and the Eastern lanes
are only weakly gravitationally bound. They are probably confined by the
external pressure of the ambient medium or by a helicoidal and/or toroidal
magnetic field (see Fiege \& Pudritz, 2000a). This is consistent with the less
efficient star-forming efficiency observed there. However, the actual mass per
unit length of the lanes is less since the O star turned on and has started to
photoevaporate the surrounding molecular gas. It is not possible to accurately
estimate the mass loss rate of both filaments as  the orientation of the dust
lanes makes it difficult to discriminate the contributions of the photoionized
lane surface layers and the nebular gas  to the free-free emission; moreover we
do not know the exact distance of the O star to the lanes. A reasonable order
of magnitude can be obtained by adopting the expression of the mass-loss rate
derived for cometary globules to the filaments, and assuming an impinging
ionizing photon flux similar to the one measured towards TC2. Under such
hypothesis, we estimate a typical mass loss rate of $85\msol\Mymu$ for a
filament $1\pc$ long. With an estimated age of $0.4\Myr$, it means that SL has
evaporated about half its initial mass, which was $\approx 85\msol\pcmu$.
Similarly EL has lost about 1/3 of its initial mass. In both cases, the initial
mass of the filament was noticeably closer to the virial mass; hence, gravity
was playing a more important role in the overall mechanical equilibrium of the
clouds, which means that the discrepancy between filament and cores with
respect to self gravity was probably much less.

Fiege \& Pudritz (2000a,b) have investigated the stability and fragmentation of
filamentary clouds truncated by an external pressure and threaded by a helical
magnetic field. They addressed the stability of the filaments to gravitational
fragmentation and axisymmetric MHD driven instabilities. The authors have
identified two well separated types of unstable modes, depending on the
toroidal to poloidal  magnetic flux ratio $\Gamma_{\Phi}/\Gamma_z$, which join
together when $\Gamma_{\Phi}/\Gamma_z= 2$. At low values, the instability mode
is gravity-driven and for pressure truncated filaments such as EL and SL, the
growth timescale is $\sim 1\Myr$ or longer. It implies that fragmentation would
have started well before the birth of the exciting star of the nebula. The
unstable MHD-driven instabilities  triggered above $\Gamma_{\Phi}/\Gamma_z = 2$
can have very large growth rates compared with gravity-driven modes, and hence
much shorter fragmentation timescales.

The regime of instability relevant to the lanes can be determined from the
ratio of the fragment separation $\lambda$ to the filament diameter $D$. In our
case, one has $\lambda/D= 1.7$  for EL and 1.5 for SL, respectively. We note
however that the photoevaporation of the outer, low-density, layers of the
filaments underestimate the actual ratio, before ionization turns on. The
radius of the filaments in the neutral parental cloud, was definitely larger,
so that the ratio is  probably closer to 1 or even less.  From Fig.~17 of FPb
it comes that $\Gamma_{\Phi}/\Gamma_z= 2-3$, i.e. MHD driven instabilities can
propagate in the two filaments. The most unstable modes are characterized by a
growth rate $\omega^2 \approx (0.003-0.1)\times (4\pi G\rho_0)^{-1}$. The
growth timescale of these modes ($\omega^{-1}$) can be as low as $4\times 10^5
\yr$ in the case ($\Gamma_{\Phi}/\Gamma_z\simeq 3$). This is of the same order
as the radial sound crossing time $R_e/c_s= 0.3\Myr$, a constraint to be met
for an instability to develop along the filament without destroying the entire
structure. We conclude that axisymmetric MHD-driven instabilities can in
principle account for the fragmentation observed in the dust lanes of the
nebula, on a timescale compatible with its age. However it puts some
constraints on the magnetic field structure of the filaments.

At this stage, our work suggests that the magnetic field probably plays an
effective role in the dynamics of the filaments of the Trifid and in their
fragmentation. Whatever the relative importance of hydrodynamical and magnetic
support, the main conclusion  is that the various condensations TC00, TC0, TC3,
TC5 most likely formed from the fragmentation of WF. Both fast and slow
fragmentation processes are at work in the Trifid, and depend on the initial
conditions in the environment of the nebula. Zeeman and polarimetric
measurements of the magnetic field are needed to quantify its role in the
support against gravity and the formation process(es) of the stellar
incubators.

\section{Conclusions}

Our mapping of the Trifid Nebula at millimeter, submillimeter  and mid-infrared
wavelengths has allowed us to characterize the gas and dust distribution.
Thirty three cores have been identified, which cover all the stages of
protostellar evolution, from the early prestellar to the late protoplanetary
phases. Sixteen cores are found to display evidence of star formation.

The cores are found to lie at various stages of photoionization, from  the
deeply embedded molecular phase to the cometary phase and the late stages of
photoevaporation. Taking into account also the previously studied objects like
TC2 (LC02), all cometary globules display very similar properties. They have
typical sizes of 0.1 to $0.2\pc$, \htwo\ densities ranging from  $\approx
10^5\cmmt$ to $10^6\cmmt$, and masses of a few \msol\ up to a few tens $\msol$.
They appear to form at most one intermediate-mass object. Hence, we conclude
that photoionization does not increase the star formation efficiency in the
individual cores. We find evidence of clustering only in the shocked, dense
condensations, such as TC3 and TC4 (previously studied in LC00) and TC5.

Comparison of the line and continuum emission properties shows hardly any
evidence for molecular depletion inside the condensations. Only moderate
molecular depletion has been reported towards the densest sources (TC3; LC00).
The mid-infrared $5-17\mu m$ flux detected with ISOCAM towards  TC1 and TC9 is
consistent with the emission of a hot central source, with dimensions typical
of protoplanetary disks, which is absorbed by the parental envelope of cold
dust and gas. ISOCAM spectroscopy has revealed the presence of large amounts of
$\rm CO_2$ ices in cometary globule TC1, thereby unveiling the potentially rich
composition of icy grain mantles in protostellar environments in the Trifid. As
the ice composition reflects the various processes that take place on grains
and the chemical evolution of the protostellar envelope, mid-infrared
observations should allow us to investigate how  the chemical composition of
the circumstellar envelopes varies along with protostellar evolution in the
\Hp\ region.

The cloud material is distributed in filamentary structures of relatively high
density (a few $10^3 \cmmt$) and gas column density $\rm N(H)$ of a few
$10^{22}\cmmd$. Their properties compare well with those of IRDCs and the Orion
filaments reported by Maddalena et al. (1986). The length of these structures
ranges from about $1\pc$ to $15\pc$ or more. Since some of them extend far
beyond the ionization front of the \Hp\ region, these filamentary structures
cannot have formed in a "collect and collapse" manner. They were pre-existing
the birth of M20 and reflect the initial conditions of the parent molecular
cloud.

Among the four filaments studied in greater detail by us, two of them are
confined by self-gravity (WF and TC4f), whereas the other two are presently
not. Taking into account the photoevaporation of the surface layers of the
latter, they were probably self-gravitating too before the onset of
photoionization. All these filaments are fragmented in condensations with
typical masses of a few hundreds of $\msol$. The fragmentation of the filaments
can be accounted for by MHD-driven instabilities, although the exact importance
of the magnetic field in the region remains to be quantified. Comparison with
numerical simulations shows that the impact of M20 on the filaments is too
recent to have triggered their fragmentation, with the exception of the TC4f
filament. It may well be, however, that this fragmentation occurred at an
earlier stage or was accelerated under the influence of high energy shock
interactions with young supernova remnants. It is not clear if the mass
distribution of the fragments in the filament WF results from the propagation
of instabilities inside the filament, hence triggering the formation of the
subsequent generations of cores. Conversely, the small-scale fragmentation
reported in the filament TC4f provides a convincing example of a filament
compressed and destabilized in the expansion of the \Hp\ region, well accounted
for by the modelling of Elmegreen (1989). The mass of the individual fragments
formed in the TC4f filament is rather low ($\sim 10\msol$), whereas the mass of
the objects in WF is typically ten times higher.

The star formation going on in the dense cores TC3 and TC5 is characterized by
very short dynamical timescales (about $10^4\yr$), comparable to the age of the
nearby W28 supernova remnant. TC5 is a good candidate to investigate the
possibility of SNR shocks as a trigger of star formation. Therefore, the
environmental conditions appear to have played a very important role in
determining  the star forming conditions in the Trifid Nebula. The kinematics
of the gas will be investigated in a forthcoming paper, in order to search for
signatures of these environmental conditions and of possible interactions with
the highly energetic phenomena which have recently taken place in the
neighbourhood.

\begin{acknowledgements}
We thank the referee for his many detailed comments, which greatly improved the
reading of the manuscript.
 J.R. Pardo and J. Cernicharo acknowledge Spanish Ministerio de Educacion y
Ciencia for supporting this research under grants AYA2003-2785, AYA2006-14876,
 ESP2004-00665, ESP2007-65812-C02-01, Communidad de Madrid for research grant
 ASTROCAM S-0505ESP-0237, and the Molecular Universe FP6 Marie Curie Research
 Training Network.
 \end{acknowledgements}

\appendix
\section{Cores in the Western Filament}

\subsection{TC5}

\begin{figure*}
\begin{center}
\leavevmode
\caption{Contour map of thermal dust emission at 1.3mm observed with the IRAM
30m telescope superposed on an $8\mu m$ image of the region (greyscale).
Contour levels are 25, 50, 75, 100, 150, 300, 450, 600, 750,900
$\mjy/11\arcsec$~beam. The Western filament WF is detected in absorption
against the background. First contour and contour interval are
$5\mjy/11\arcsec$ beam. A magnified view of the protostellar cluster associated
with the TC5 core is presented in the right panel. } \label{iram1300_TC5}
\end{center}
\end{figure*}
\smallskip

TC5 is the brightest source discovered in our survey, with a peak flux of
$\simeq 1Jy/11\arcsec$ beam. It is located in  the southernmost fragment
($\approx 810\msol$) of the Western filament WF
(Figs.~\ref{iramsurvey}-\ref{iram1300_TC5}). The isocontours of the 20cm
free-free emission show that the core lies at the edge of the W28 supernova
remnant.

The core is elliptical with major and minor axis of $0.20\pc \times 0.13\pc$,
respectively. The brightness peak is off-centered, suggesting the presence of
multiple sources inside the continuum core. Indeed, SPITZER unveiled  a cluster
of sources in the dust central core (see right panel in
Fig.~\ref{iram1300_TC5}). The emission is strongly concentrated towards the
peak. A mass of $100\msol$ (about half the core mass) is contained within a
$9\arcsec$ region, which indicates \htwo\ densities $\rm n(\htwo)\simeq 8\times
10^6\cmmt$).  The continuum source is not detected in the near-infrared with
2MASS. It however coincides with the IRAS pointsource 17589-2312. Comparison of
the bolometric luminosity ($4000\lsol$) to the millimeter luminosity yields a
ratio $\rm L_{bol}/(10^3\times L_{1.25})\simeq 15$. This value is low and
typical of those encountered in Class 0 protostellar cores. This testifies to
the youth of the TC5 core.

The line profiles display broad wings tracing gas that reach velocity of
$20\kms$ with respect to the  core, which testifies to the ongoing star
formation at work in the core (Fig.~\ref{panel_line_TC5}). The discovery of
methanol maser emission confirms indeed that massive protostars or non-ionizing
intermediate-mass protostars are forming in TC5 (Pestalozzi et al. 2005).
Complementary interferometric observations are required to identify which of
the continuum sources are powering the molecular outflows detected towards the
core (Lefloch et al. 2008). The physical properties of TC5 are therefore very
similar to those of  TC3 and TC4.  At the high densities encountered in the
central core of TC5, the typical gravitational free-fall time is $\tau=
(3\pi/32G\rho)^{1/2}\simeq 1.2\times 10^4\yr$. The entire star-forming process
taking place in the core is very recent.

\begin{figure}
\begin{center}
\leavevmode
\caption{Molecular gas emission obtained towards TC5 in the millimeter
transitions of SiO and CS. Fluxes are expressed in units of antenna
temperature. } \label{panel_line_TC5}
\end{center}
\end{figure}
\smallskip

\subsection{TC6}

The thermal dust emission shows a weakly contrasted condensation, without any
dominant central condensation (Fig.~\ref{iram_TC6}). Instead, there are several
components barely resolved by the telescope beam, separated by typically
$20\arcsec$. These sub-condensations (TC6A-D) have low masses of typically
$2-3\msol$.

Only a few molecular lines were observed towards TC6, at SEST and CSO. As the
telescope main beam has a size comparable to the dimensions of the
condensations, the main-beam filling factor  must be taken into account to
accurately estimate the brightness of the CS lines and the density of the
condensations. The filling factor $f$ is  0.81 and 0.65 for the \jthreetotwo\
and the \jtwotoone\ transitions, respectively, assuming a Gaussian source
distribution $f= \theta_s^2/(\theta_s^2+\theta_B^2)$. Hence,  the CS line
brightness temperature of the \jthreetotwo\ line components are $0.49\K$ and
$0.35\K$, respectively, and $0.74\K$ and $0.65\K$ for the \jtwotoone\ line.

Neither ISO nor SPITZER detected pointlike sources towards TC6, which is
consistent with TC6 being in a prestellar phase (see Fig.~\ref{iramsurvey}).
Indeed, the mass of the subcondensations is much less than the virial
equilibrium mass (see Sect.~7).  Two kinematic components are detected in the
CS lines (see Table~5). The line profiles are broader than those observed in
the dust lanes of the nebula, and indicate a higher degree of turbulence.

\begin{figure}
\begin{center}
\leavevmode
\caption{ Thermal dust emission map of TC6 at 1.3mm observed with the IRAM 30m
telescope. First contour and contour interval are $5\mjy/11\arcsec$
beam.}\label{iram_TC6}
\end{center}
\end{figure}
\smallskip

\begin{figure}
\begin{center}
\leavevmode
\caption{ Millimeter line emission observed at SEST in TC6 at the offset
position $(-280\arcsec,-440\arcsec)$. The CS line intensities have been
corrected for main beam dilution effects.} \label{panel_line_TC6}
\end{center}
\end{figure}
\smallskip

\section{Cores in the Eastern Lane}

\subsection{TC8}

Figure~\ref{iram_TC1} shows a second condensation  gas in the Eastern lane.
Emission of the dense layers gas was observed in CS in the \jtwotoone\
transition at IRAM and in the \jthreetotwo\ transition at SEST. Unfortunately,
the fragment and TC8 could not be fully mapped.

The molecular emission peaks near the offset position
($+93.6\arcsec$,$+88.6\arcsec$) where a Class I protostar has been discovered
by SPITZER (Rho et al. 2006). The lines of CO and its isotopologues have
intensities comparable to those measured in TC1 (Fig.~\ref{panel_line_tc1}),
indicating a kinetic temperature of about $30\K$. Due to incomplete mapping, we
have estimated the physical conditions in the core from the emission at the
position ($+90\arcsec$, $+80\arcsec$) close to the SPITZER continuum source.
Because the core size is of the same order as the telescope beam size, the
effects of the convolution with the main beam cannot be neglected, unlike the
case of TC1, whose extent is typically  twice the telescope beam size. The
intensity of the CO \jtwotoone\ line is $\approx 0.75\K$ at the brightness
peak. The beam filling factor reads $\theta_s^2/(\theta_s^2+\theta_B^2)$ where
$\theta_s= 20\arcsec$ and $\theta_B= 25.3\arcsec$ are the source and beam size
(HPFW), respectively. A good estimate of the line brightness is obtained by
dividing the main-beam brightness temperature by the beam filling factor. In
the case of the CS \jtwotoone\ line, the beam filling factor is $\approx
0.385$, and the brightness temperature of the line is $1.9\K$. From the
main-beam brightness temperature measured at SEST, we estimate a brightness
temperature of $2.2\K$ for the \jthreetotwo\ line (the beam size is
$33.7\arcsec$ at this frequency). Hence, the line intensities compare well with
those measured towards TC1.

As a conclusion, TC8 is characterized by molecular gas and dust emission which
is less constrated than in TC1. However, the mass and the physical conditions
in both condensations are rather similar. The dense gas region in TC8, as
traced by CS, occupies a smaller part of the molecular gas fragment. The mass
of the CS core represents only 25\% of the fragment mass associated with TC8.
In TC1, the CS core represents 80\% of the whole mass. We interpret this
difference as a result of a more effective photoionization and compression of
TC1.

\subsection{TC13}

\begin{figure}
  \begin{center}
    \leavevmode
\caption{ Molecular line spectra obtained at the emission peak of the molecular
core TC13. Fluxes are expressed in units of antenna temperature. The dashed
line marks the velocity of the main body gas.} \label{panel_line_TC13}
  \end{center}
\end{figure}
\smallskip

The Eastern lane bends in the Southeastern direction beyond TC8 (see
Fig.~\ref{iramsurvey}). This part of the dark lane contains lower amounts of
gas and dust. As a consequence, it is detected in molecular gas emission (CO
isotopomers and CS) while it is missed by our 1.3~mm continuum mapping. A
condensation of $55\arcsec \times 25\arcsec$ was discovered in this part of the
filament (Fig.~\ref{iram_tc13}).  A Class~0/I source candidate (TC13) was
identified by SPITZER, which lies in the outer layers of the condensation.

The CS \jtwotoone\ transition was detected at SEST. The line has a typical
main-beam brightness intensity $\rm T_{mb}^{21}= 0.6\K$. Once corrected for
beam dilution (the beam size is comparable to the size of the core), we
estimate a brightness temperature $\rm T^{21}_{B}= 1.3\K$ at the brightness
peak (Fig.~\ref{panel_line_TC13}). Assuming a standard relative abundance $\rm
[CS]/[\htwo]= 10^{-9}$, the CS column density is $\rm N(CS)\sim 6\times
10^{12}\cmmd$. The gas density must be $\simeq 10^5\cmmt$ in order to account
for the brightness of the CS \jtwotoone\ line.

\begin{figure}
  \begin{center}
    \leavevmode
\caption{Molecular gas emission observed towards TC13, integrated between 2 and
$8\kms$. The arrow points to the position of the protostellar
source.}\label{iram_tc13}
  \end{center}
\end{figure}
\smallskip

\section{Cores in the Southern Region}

\begin{figure}
  \begin{center}
    \leavevmode
\caption{Molecular line emission in the Southern Lane region, integrated
between -3 and $+3\kms$, showing TC10 and TC11.} \label{map_tc10}
\end{center}
\end{figure}
\smallskip

\begin{figure}
  \begin{center}
    \leavevmode
\caption{ Molecular line spectra observed at the emission peak of  TC10 in the
Southern Lane. Fluxes are expressed in units of antenna
temperature.}\label{panel_tc10}
  \end{center}
\end{figure}
\smallskip

Two condensations, TC10 and TC11 respectively, could be identified in our
molecular emission maps (see Fig.~\ref{map_tc10}). The two condensations are
separated by $\approx 84\arcsec= 0.7\pc$. Only the first condensation is
detected in the millimeter continuum. The overall emission is weak; the peak
flux is about $25 \mjy/11\arcsec$ beam, and the integrated flux is $25\mjy$,
implying a peak column density $\rm N(H)= 2.3\times 10^{22}\cmmd$. The central
core, defined by the contour at half power, has a typical size of $7\arcsec$
(beam-deconvolved) and a mass of $2.5\msol$.

The molecular gas condensation, defined by the contour at half-power, has an
elliptical shape with major (minor) axis of $80\arcsec$ ($30\arcsec$) contains
a mass of $25\msol$, almost one third  of the material contained in the lane
($66\msol$).

Lines are somewhat narrower in the Southern lane. We measure $\rm T_{mb}^{21}=
2.5\K$ ($\Delta v= 1.4\kms$) (Fig.~\ref{panel_tc10}). Assuming a standard
relative abundance $\rm [CS]/[\htwo]= 10^{-9}$, the CS line emission is
consistent with \htwo\ densities of $2.5\times 10^5\cmmt$. In the Northern
region, the molecular gas emission reveals TC11 as a larger condensation of
larger dimensions and lower gas column density. Adopting the same physical
conditions as in the South, we derive a peak column density $\rm N(\htwo)=
9\times 10^{21}\cmmd$ and a mass of $6\msol$ from the \thco\ data. No evidence
of protostellar activity was detected in the Southern lane. This is consistent
with the mass of TC10 and TC11 being much lower than their virial mass.


\begin{thebibliography}{}
\bibitem[]{andre} Andr\'e, P., Ward-Thompson, D., Barsony, M., 2000,
  in Protostar \& Planets IV,
\bibitem[]{} Bacmann, A., Lefloch, B., Ceccarelli, C., et al., 2002, A\&A, 389,
L6
\bibitem[]{} Bertoldi, F., 1989, \apj, 346, 735
\bibitem[]{} Beuther, H., Churchwell, E., McKee, C.F., Tan, J.C., 2007,
in {\em Protostars and Planets V}, eds. B. Reipurth, D. Jewitt, \& K. Keil
(Tucson~: Univ. Arizona Press)
\bibitem[]{} Broguiere, D., Neri, R., Sievers, A., 1995, NIC bolometer users
guide (IRAM internal report)
\bibitem[]{} Cernicharo, J., Lefloch, B., Cox, P., et al., 1998,
Science, 282, 462 (CL98)
\bibitem[]{}  Cernicharo, J., Noriega-Crespo, A., Cesarsky, D., et al.,
2000, Science,  288, 649
315, L32
  Rouan, D., 1999, \aap, 352, 277
\aap, 315, L49

\bibitem[]{} Dobashi, K., Yonekura, Y., Matsumoto, T., et al., 2001, PASP, 53,
85
\bibitem[]{} Deharveng, L., Lefloch, B., Massi, F., et al., 2006, \aap,
458, 191
\bibitem[]{} Deharveng, L., Lefloch, B., Zavagno, A., et al,. 2003, \aap,
408, L25
  291, 561
\bibitem[]{} Dobashi, K., Yonehura, Y., Matsumoto, T., et al., 2001,
PASJ, 53,85
\bibitem[]{} Elmegreen, B.G., Lada, C.J., 1977, \apj, 214, 725
\bibitem[]{} Elmegreen, B.G., 1989, \apj, 340, 786
\bibitem[]{} Elmegreen, B.G., 2002, ApJ, 577, 206
\bibitem[]{} Fiege, J.D., Pudritz, R.E., 2000a, MNRAS, 311, 85
\bibitem[]{} Fiege, J.D., Pudritz, R.E., 2000b, MNRAS, 311, 105
\bibitem[]{} Fukuda, N., Hanawa, T., 2000, \apj, 533, 911
\bibitem[]{} Healy, K.R.,Hester,J. M., Claussen, M.J., 2004, \apj, 610, 835
\bibitem[]{} Hester, J.J., Scowen, P.A., Sankrit, R., 1996, \aj, 111, 2349
  \apj, 377, 192
  Luhmna, M.L., 1999, \apj, 527, 795 (K99)
315, L27
\bibitem[]{} Lefloch, B., Lazareff, B., 1994, \aap, 289, 559 (LL94)
\bibitem[]{} Lefloch, B., Lazareff, B., 1995, \aap, 301, 522
\bibitem[]{blf1} Lefloch, B., Lazareff, B., Castets, A., 1997, \aap,
  324, 249
\bibitem[]{} Lefloch, B., Cernicharo, J., 2000, \apj, 545, 340 (LC00)
\bibitem[]{blf2} Lefloch, B., Cernicharo, J., Cesarsky, D., et al., 2001,
\aap, 368, L13 (L01)
\bibitem[]{} Lefloch, B., Cernicharo, J., Rodriguez, L.F., et al., 2002,
\apj, 581, 335 (L02)
\bibitem[]{} lefloch, B., Cernicharo J., 2003, in The Young Local Universe,
proceedings. of the XXXIXth Rencontres de Moriond, eds. A. Chalabaev, T.
Montmerle, J. Tran Thanh Van, pg 353,
\bibitem[]{} Lynds, B., Canzian, B.J., and O'Neil, E.J. Jr.,
  1985, \apj, 288, 164
\bibitem[]{} Lynds, B., O'Neil, E.J. Jr., 1986, AJ, 92, 1125
\bibitem[]{} Martens, F., Schaerer, D., Hillier, D.J., 2005, A\&A, 436, 1049
\bibitem[]{} Maddalena, R.J., Morris, M., Moscowitz, J., et al., 1986, \apj,
303, 375
\bibitem[]{} Minier, V., Burton, M.G., Hill, T., et al. 2005, A\&A, 429, 945
\bibitem[]{} Motte, F., Andr\'e, P., Neri, R., 1998, A\&A, 336, 150
\bibitem[]{} Pestalozzi, M.R., Minier, V., Booth, R.S., 2005, A\&A, 432, 737
\bibitem[]{} Nakamura,F.,  Hanawa, T., Nakano, T., 1993, PASJ, 45 551
\bibitem[]{} Panagia, N., 1973, AJ, 78, 929
\bibitem[]{rei83} Reipurth, B., 1983, \aap, 117, 183
\bibitem[]{rei98} Reipurth, B., Bally, B., Fesen, R.A., Devine, D., 1998,
  Nature, 396, 343
\bibitem[] Rho, J., Borkowski, K.J., 2002, \apj, 575, 201
\bibitem[]{} Rho, J., Corcoran, M.F., Chu, Y.H., Reach, W.T., 2001, \apj,
  562, 446
\bibitem[]{} Rho, J., Reach, W., Lefloch, B., Fazio, G., 2006, \apj, 643,965
\bibitem[]{} Rosado, M., Esteban. C., Lefloch, B., Cernicharo, J.,
  Garc\'{\i}a-L\'opez, R.J., 1999, AJ, 118, 2962
  \apj, 374, 564
  Erickson, E.F., Haas, M.R., 1994, \apj, 420, 772
  Haas, M.R., 1995, \apj, 444, 721
\bibitem[]{} Spitzer, L., 1978, {\em Physical Processes in the Interstellar
    Medium}, (New-York~: Wiley), 107
\bibitem[]{sugi1} Sugitani, K., Fukui, Y., Ogura, K., 1991, \apj Supp, 77, 59
\bibitem[]{} Tomisaka, K., 1995, \apj, 328, 226
Caux, E., 2001, \aap, 376, 1064
  Wright, C.M., Timmermann, R., 1996, \aap, 315, 337
\bibitem[]{} Vel\'azquez, P.F., Dubner, G.M., Goss, W.M., Green, A.J., 2002,
AJ, 124, 2145
\bibitem[]{} Walborn, N.R., 1973, \aj, 78, 1067
\bibitem[]{} Walmsley,C.M., Pineau des For\^ets, G., Flower, D.R., 1999,
  \aap, 342, 542
\bibitem[]{} White, G.J., Nelson, R.P., Holland, W.S., et al., 1999, \aap,
  342, 233
\bibitem[]{} Whitworth, A.P., Bhattal, A.S., Chapman, S.J., 1994, MNRAS,
  268, 291
344, 770
358, 116
\bibitem[]{} Yamaguchi,R., Saito, H., Mizuno, N., et al., 1999, PASJ, 51, 791
\bibitem[]{} Yusef-Zadeh, F., Shure,M., Wardle, M., Kassim, N., 2000, \apj,
540, 842
\bibitem[]{} Yusef-Zadeh, F., Biretta J., Geballe, T.R., 2005, AJ, 130, 1171
\bibitem[]{} Zavagno, A., Pomares, M., Deharveng, L., et al., 2007,\aap, 472,
835
\end{thebibliography}
\end{document}